\documentclass[12pt]{article}
\usepackage{a4wide}
\usepackage{latexsym}
\usepackage{amsmath}
\usepackage{amsfonts}
\usepackage{amscd}
\usepackage{amsthm}
\usepackage{shuffle}
\usepackage{cite}
\usepackage{graphicx}
\usepackage{float}
\usepackage{axodraw}

\usepackage{pslatex}
\usepackage[latin1,utf8]{inputenc}
\usepackage[OT2,T1]{fontenc}

\usepackage{ifthen}

\allowdisplaybreaks

\newcommand{\bq}{\begin{eqnarray}}
\newcommand{\eq}{\end{eqnarray}}
\newcommand{\eps}{\varepsilon}

\newcommand{\loopnumber}{l}
\newcommand{\coupling}{g}

\usepackage{color}
\newcommand{\sw}{}

\begin{document}

\thispagestyle{empty}

\begin{flushright}
  MITP/19-040
\end{flushright}

\vspace{1.5cm}

\begin{center}
  {\Large\bf Integrands of loop amplitudes within loop-tree duality \\
  }
  \vspace{1cm}
  {\large Robert Runkel, Zolt\'an Sz\H{o}r, Juan Pablo Vesga and Stefan Weinzierl\\
\vspace{2mm}
      {\small \em PRISMA Cluster of Excellence, Institut f{\"u}r Physik, }\\
      {\small \em Johannes Gutenberg-Universit{\"a}t Mainz,}\\
      {\small \em D - 55099 Mainz, Germany}\\
  } 
\end{center}

\vspace{2cm}

\begin{abstract}\noindent
  {
Using loop-tree duality, we relate a 
renormalised $n$-point $\loopnumber$-loop amplitude in a quantum field theory 
to a phase-space integral
of {\sw a regularised $\loopnumber$-fold forward limit of a 
UV-subtracted 
$(n+2\loopnumber)$-point tree-amplitude-like object.
We show that up to three loops the latter object is easily computable from recurrence relations.
}
This defines an integrand of the loop amplitude with a global definition of the loop momenta.
Field and mass renormalisation are performed in the on-shell scheme.
   }
\end{abstract}

\vspace*{\fill}

\newpage

\section{Introduction}
\label{sect:intro}

Precision physics at the LHC requires next-to-next-to-leading order (NNLO) calculations for various processes.
If one goes beyond the simplest $2 \rightarrow 2$-processes towards
$2 \rightarrow (n-2)$-process with $n>4$, one is in particular interested
in efficient methods which allow for automation.
Numerical methods 
like numerical loop integration \cite{Soper:1998ye,Soper:1999xk,Nagy:2003qn,Gong:2008ww,Assadsolimani:2009cz,Assadsolimani:2010ka,Becker:2010ng,Becker:2011vg,Becker:2012aq,Becker:2012nk,Becker:2012bi,Goetz:2014lla,Seth:2016hmv,Anastasiou:2018rib}
combined with loop-tree duality \cite{Catani:2008xa,Bierenbaum:2010cy,Bierenbaum:2012th,Buchta:2014dfa,Hernandez-Pinto:2015ysa,Buchta:2015wna,Sborlini:2016gbr,Driencourt-Mangin:2017gop,Driencourt-Mangin:2019aix,Aguilera-Verdugo:2019kbz,Runkel:2019yrs,Baumeister:2019rmh,Capatti:2019ypt,Capatti:2019edf}
or methods based on numerical unitarity {\sw \cite{Gluza:2010ws,Kosower:2011ty,CaronHuot:2012ab,Zhang:2012ce,Sogaard:2014ila,Larsen:2015ped,Ita:2015tya,vonManteuffel:2014ixa,Peraro:2016wsq,Peraro:2019svx,Badger:2017jhb,Badger:2018gip,Badger:2018enw,Abreu:2017idw,Abreu:2017xsl,Abreu:2017hqn,Abreu:2018jgq,Abreu:2018zmy,Abreu:2019odu} }
are a promising path for this approach.
Let us mention that there is in addition the $Q$-cut approach \cite{Baadsgaard:2015twa,Huang:2015cwh,Feng:2016msc}, sharing some similarities with
loop-tree duality, but differing in the details.

In this paper we focus on the loop-tree duality approach.
We show that the integrand for the renormalised $n$-point $\loopnumber$-loop amplitude 
within the loop-tree duality approach 
is related to the regularised $\loopnumber$-fold forward limit of 
{\sw a UV-subtracted $(n+2\loopnumber)$-point tree-amplitude-like object}, 
if field renormalisation and mass renormalisation are performed in the on-shell scheme.

The applications of our result are twofold:
First of all, it is a significant efficiency improvement. We are no longer forced to consider individual Feynman
diagrams, whose number grows drastically with the number of external legs and the number of loops.
Instead, we may entirely work at the integrand level with tree-amplitude-like objects.
We remind the reader that tree amplitudes may be computed numerically in an efficient way through 
recursion relations \cite{Berends:1987me}.
For example, the CPU cost for a cyclic-ordered tree amplitude with $n$ external particles in a quantum field
theory with cubic vertices scales as $n^3$.

Secondly, the infrared limits of tree amplitudes are very well understood. 
We know where they are (in the region where internal propagators go on-shell) and how to compute
the limiting behaviour. 
This may now be transferred to the tree-amplitude-like objects. 
Our formulation is a further step
towards a local cancellation of infrared divergences between real and virtual contributions.
In particular, our result provides naturally a global definition of the loop momenta \cite{Tourkine:2019ukp}.

We present the equivalence between the renormalised $n$-point $\loopnumber$-loop amplitude 
and the phase space integral of 
{\sw a regularised $\loopnumber$-fold forward limit of a UV-subtracted $(n+2\loopnumber)$-point tree-ampli\-tude-like object}
as a general property of quantum field theory.
For the field renormalisation and the mass renormalisation we use the on-shell scheme (for reasons which will become clear
in the main part of the article).
For the renormalisation of the coupling and any other quantities we may take any renormalisation scheme.
In addition, one easily obtains results where the field renormalisation or the mass renormalisation are performed in a
scheme different from the on-shell scheme through a (ultraviolet-) finite renormalisation.

In order to keep the notation to a minimum, we focus on theories, where all fields have a vanishing vacuum expectation
value.
This includes theories like Yang-Mills theory and QCD, but not electroweak theory, where the Higgs field
has a non-vanishing expectation value.
The extension towards fields with non-vanishing vacuum expectation values is straightforward, and we indicate
the necessary steps in a dedicated section.

Integrands for loop amplitudes for specific theories (possibly with additional restrictions on the number of loops or planarity)
have been considered in the literature before \cite{ArkaniHamed:2010kv,ArkaniHamed:2012nw,Arkani-Hamed:2013jha,Arkani-Hamed:2013kca,Geyer:2015bja,Geyer:2015jch,Geyer:2016wjx,Geyer:2017ela,Geyer:2018xwu,Cachazo:2015aol,Feng:2016nrf,He:2016mzd,He:2017spx,Salvatori:2018fjp,Salvatori:2018aha}.
In this article we present a general result, 
{\sw which expresses the integrand of a renormalised $n$-point $l$-loop amplitude as a regularised $\loopnumber$-fold forward limit of a 
UV-subtracted $(n+2\loopnumber)$-point tree-amplitude-like object, which is valid in a generic quantum field theory
without any restrictions on the number of loops.
We also discuss how to compute these objects efficiently from recurrence relations. Here we limit ourselves to three loops or less.}

Ideas of relating loop integrands to the forward limit have appeared before 
in the literature \cite{CaronHuot:2010zt,He:2015yua}.
In this paper we address all technical challenges associated to this approach.
The technical challenges are related to the fact, that in general the forward limit of tree amplitudes is singular.
This is due to internal propagators, which go on-shell in the forward limit.
Let us first note that the forward limit of a tree diagram is equivalent to the integrand of a loop diagram with cut propagators.
Each forward limit of a pair of external lines of a tree diagram corresponds to a cut loop propagator.
We may group the singular configurations into three categories:
The first category consists of diagrams, which correspond to self-energy corrections on external lines.
The Lehmann-Symanzyk-Zimmermann (LSZ) reduction formula \cite{Lehmann:1954rq} 
instructs us to omit these.
The second category consists of diagrams, which contain tadpole sub-diagrams, connected to the rest of the diagram by a massless line.
In theories, where the field has a vanishing vacuum expectation value, these diagrams cancel with counterterms coming from
the renormalisation of the source. 
As the sum of the two contributions vanishes, it is common practice to omit both contributions.
The third category consists of diagrams, which correspond to self-energy corrections on internal lines.
On the loop side, they correspond to diagrams with higher powers of some propagators.
Quite recently, it was shown that these contributions cancel with similar contributions from the ultraviolet counterterms
for the field and the mass
in the on-shell scheme \cite{Baumeister:2019rmh}.
For this reason we consider renormalised loop amplitudes, where the field and the masses are renormalised in the on-shell scheme.
We define the regularised $\loopnumber$-fold forward limit of a tree amplitude as the limit, where the singular contributions
have been subtracted out.

In addition, there are some non-trivial combinatorial issues:
Loop diagrams come with symmetry factors and additional minus signs for closed fermion loops. 
{\sw Loop-tree duality introduces additional combinatorial factors.
We show how the symmetry factors of loop diagrams are matched on the tree-diagram side.}
Minus signs for closed fermion loops have a correspondence with minus signs appearing when a tree amplitude with identical
fermion pairs is expressed in terms of tree amplitudes with non-identical fermions.
{\sw We discuss in detail how the additional combinatorial factors from the application of loop-tree duality are incorporated.}

The main result of this paper is given in {\sw eq.~(\ref{main_result}),
which relates
the renormalised $n$-point $\loopnumber$-loop amplitude 
to a phase space integral of a regularised $\loopnumber$-fold forward limit of a UV-subtracted $(n+2\loopnumber)$-point tree-amplitude-like object}.
Let us stress, that in order to obtain this simple result we must ensure that contributions from residues
underlying higher poles vanish.
This happens if the fields and the masses are renormalised in the on-shell scheme.
Our result would not be as simple if we would consider the unrenormalised amplitude or a renormalised amplitude
with masses renormalised in the $\overline{\mathrm{MS}}$-scheme.
However, let us point out that our result is also useful if one is interested in renormalised loop amplitudes
with masses renormalised in the $\overline{\mathrm{MS}}$-scheme (or any other mass renormalisation scheme):
One may always compute first the result with masses renormalised in the on-shell scheme and then perform
a finite renormalisation to switch to the desired mass renormalisation scheme.
The steps required for the latter calculation are usually simpler than for a full calculation.

This paper is organised as follows:
In the next section we introduce the basic notation.
In section~\ref{sect:graphs} we define various sets of graphs, which will be relevant to our discussion.
In section~\ref{sect:cutting_sewing} we introduce two operations on graphs: Cutting and sewing, which are at the level of graphs
inverse to each other.
In section~\ref{sect:loop_amplitudes} we review the essential features of loop-tree duality.
The next three sections are devoted to the technical challenges:
We define the regularised forward limit of a tree amplitude in section~\ref{sect:forward_limit},
discuss symmetry factors of loop diagrams in section~\ref{sect:symmetry_factors}
and review the absence of contributions from higher poles in the on-shell scheme in section~\ref{sect:higher_powers}.
After this preparation, we present our main result -- the equivalence of the loop integrand within the loop-tree duality approach with the
regularised forward limit -- in section~\ref{sect:integrand}.
Section~\ref{sect:recurrence_relations} is of practical nature and discusses the efficient computation of the integrand
with the help of recurrence relations.
Section~\ref{sect:vev} sketches the required steps for a generalisation of our result towards theories with fields with
non-vanishing vacuum expectation values.
{\sw In section~\ref{sect:checks} we report on the checks we performed.}
Finally, section~\ref{sect:conclusions} contains our conclusions.

\section{Basic notation}
\label{sect:basic_notation}

We consider loop amplitudes in a quantum field theory.
It could be a scalar theory, Yang-Mills theory, QCD, etc..
For illustration purposes we will often choose one of the simplest quantum field theories,
a scalar $\phi^3$-theory with generating functional
\bq
 Z\left[J\right] 
 & = & 
 \int {\mathcal D} \phi(x) \; 
  \exp\left(i \int d^Dx \; {\mathcal L} \right)
\eq
and Lagrangian
\bq
\label{counterterms}
 {\mathcal L} 
 & = &
 \frac{1}{2} \left( \partial_\mu \phi \right) \left( \partial^\mu \phi \right)
 - \frac{1}{2} m^2 \phi^2
 + \frac{1}{3!} \coupling^{(D)} \phi^3
 + J \phi
 +  {\mathcal L}_{\mathrm{CT}}.
\eq
\bq
 {\mathcal L}_{\mathrm{CT}}
 & = &
 - \frac{1}{2} \left(Z_\phi-1\right) \phi \Box \phi 
 - \frac{1}{2} \left(Z_\phi Z_m^2 -1 \right) m^2 \phi^2
 + \frac{1}{3!} \left(Z_\phi^{\frac{3}{2}} Z_\coupling - 1 \right) \coupling^{(D)} \phi^3
 + \left(Z_\phi^{\frac{1}{2}} Z_J - 1 \right) J \phi.
 \nonumber \\
\eq
We have written the Lagrangian in renormalised quantities. 
Eq.~(\ref{counterterms}) gives the appropriate ultra\-violet counterterms.
Please note that we included the renormalisation of the source $J$.

Within dimensional regularisation it is convenient to relate
the bare coupling $\coupling_{\mathrm{bare}}$ to the renormalised coupling $\coupling$
by
\bq
 \coupling_{\mathrm{bare}} & = & Z_\coupling S_\eps^{-\frac{1}{2}} \mu^\eps \coupling,
\eq
where the quantity
\bq
 S_\eps 
 & = & 
 \left( 4 \pi \right)^\eps e^{-\eps\gamma_E}
\eq
is the typical phase space volume factor in $D =4-2\eps$ dimensions and
$\gamma_E$ is Euler's constant.
The additional factor $\mu^\eps$ keeps the mass dimension of the renormalised coupling $\coupling$ constant.
$\coupling^{(D)}$ is given by
\bq
 \coupling^{(D)} & = & S_\eps^{-\frac{1}{2}} \mu^\eps \coupling.
\eq
Let us consider an amplitude with $n$ external particles in a given quantum field theory.
We denote the coupling by $\coupling$, as we did in the example of $\phi^3$-theory.
We take all external momenta to be outgoing. The external momenta are on-shell
\bq
 p_i^2 & = & m_i^2
\eq
and satisfy momentum conservation
\bq
 p_1 + p_2 + ... + p_n & = & 0.
\eq
We denote the renormalised $\loopnumber$-loop amplitude with $n$ external particles by
\bq
 {\mathcal A}_{\loopnumber,n}\left(p_1,...,p_n\right).
\eq
If we consider particles with spin, we denote by $\lambda_i$ their helicities and write
\bq
 {\mathcal A}_{\loopnumber,n}\left(p_1^{\lambda_1},...,p_n^{\lambda_n}\right).
\eq
If the quantum field theory under consideration contains different particle species, we introduce an additional label
which distinguishes the different species.

The amplitude ${\mathcal A}_{\loopnumber,n}$ is of order 
\bq
 \coupling^{2\loopnumber+n-2}
\eq
in the coupling.

The relation between the renormalised and the bare amplitude (for the example of $\phi^3$-theory) is given 
to all orders in $\coupling$ by
\bq
\label{LSZ}
 {\mathcal A}(p_1,...,p_n,g,m)
 & = & 
 \left( Z_\phi^{1/2} \right)^{n}
 {\mathcal A}^{\mathrm{bare}}(p_1,...,p_n,g_{\mathrm{bare}},m_{\mathrm{bare}})
 \nonumber \\
 & = &
 \left( Z_\phi^{1/2} \right)^{n}
 {\mathcal A}^{\mathrm{bare}}\left(p_1,...,p_n,Z_g S_\eps^{-\frac{1}{2}} \mu^\eps g, Z_m m \right).
\eq
Let us now expand eq.~(\ref{LSZ}) in the coupling. We write
\bq
 {\mathcal A}_{\loopnumber,n}(p_1,...,p_n,g,m)
 & = &
 {\mathcal A}_{\loopnumber,n}^{\mathrm{bare}}(p_1,...,p_n,g,m) + {\mathcal A}_{\loopnumber,n}^{\mathrm{CT}}(p_1,...,p_n,g,m),
\eq
where both expressions on the right-hand side are expressed in terms of renormalised quantities and
${\mathcal A}_{\loopnumber,n}^{\mathrm{CT}}$ contains exactly all contributions from ultraviolet counterterms.

The renormalised amplitude ${\mathcal A}_{\loopnumber,n}$ depends on the chosen renormalisation scheme. 
If we change the renormalisation scheme from a scheme $\mathrm{CT}$ to a scheme $\mathrm{CT}'$ we have
\bq
 {\mathcal A}_{\loopnumber,n}'(p_1,...,p_n,g',m')
 & = &
 {\mathcal A}_{\loopnumber,n}^{\mathrm{bare}}(p_1,...,p_n,g',m') + {\mathcal A}_{\loopnumber,n}^{\mathrm{CT}'}(p_1,...,p_n,g',m').
\eq
On the other hand we may relate ${\mathcal A}_{\loopnumber,n}'$ and ${\mathcal A}_{\loopnumber,n}$:
\bq
\label{finite_renorm}
 {\mathcal A}_{\loopnumber,n}'(p_1,...,p_n,g',m')
 & = &
 {\mathcal A}_{\loopnumber,n}(p_1,...,p_n,g,m) + \Delta {\mathcal A}_{\loopnumber,n}^{\mathrm{finite}}(p_1,...,p_n,g,m),
\eq
where $\Delta {\mathcal A}_{\loopnumber,n}^{\mathrm{finite}}$ describes the change due to the (ultraviolet-) finite renormalisation.
Eq.~(\ref{finite_renorm}) allows us to compute the renormalised amplitude ${\mathcal A}_{\loopnumber,n}$ first in one
renormalisation scheme $\mathrm{CT}$ (which we may choose based on technical advantages)
and then transfer the result to the desired renormalisation scheme
$\mathrm{CT}'$.
The (ultraviolet-) finite renormalisation from $\mathrm{CT}$ to $\mathrm{CT}'$ is in general simpler than the calculation
of the renormalised loop amplitude ${\mathcal A}_{\loopnumber,n}$.
In this paper we focus on the calculation of the renormalised loop amplitude ${\mathcal A}_{\loopnumber,n}$.

Let us comment on gauge theories: It is well-known that in gauge theories the field renormalisation constants are in general
gauge-dependent and cancel the gauge-dependence of ${\mathcal A}^{\mathrm{bare}}$.
The renormalised amplitude ${\mathcal A}(p_1,...,p_n,g,m)$ is of course gauge-independent \cite{Boehm}.

\section{Graphs}
\label{sect:graphs}

\begin{figure}
\begin{center}
\includegraphics[scale=1.0]{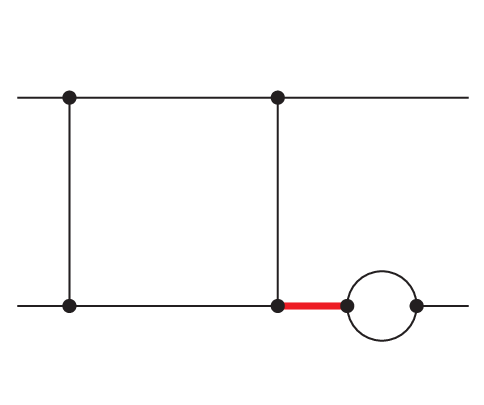}
\end{center}
\caption{
A graph with a self-energy insertion on an external line.
This graph belongs to ${\mathcal U}_{2,4}$, but not to ${\mathcal U}_{2,4}^{\mathrm{amputated}}$.
The graph contains an on-shell propagator, indicated by a red line.
}
\label{fig_external_self_energy}
\end{figure}

Let us denote the set of all (unordered) connected graphs with $n$ external edges 
of order $\coupling^{2\loopnumber+n-2}$ in the coupling
by
\bq
 {\mathcal U}_{\loopnumber,n}.
\eq
The set ${\mathcal U}_{\loopnumber,n}$ contains all connected graphs with $n$ external edges and $\loopnumber$ loops,
which can be drawn according to the Feynman rules. The set ${\mathcal U}_{\loopnumber,n}$ also 
includes graphs with ultraviolet counterterms.
For $\loopnumber \ge 1$ the set ${\mathcal U}_{\loopnumber,n}$ contains graphs corresponding to self-energy corrections on external lines.
An example is shown in fig.~\ref{fig_external_self_energy}.
We denote the set of graphs without self-energy corrections on external lines by
\bq
 {\mathcal U}_{\loopnumber,n}^{\mathrm{amputated}}.
\eq 
Graph with self-energy corrections on external lines are problematic, because they contain an internal on-shell propagator, 
indicated by a red line in fig.~\ref{fig_external_self_energy}.
Basically, we would divide by zero.
Luckily, the Lehmann-Symanzyk-Zimmermann (LSZ) reduction formula \cite{Lehmann:1954rq} 
instructs us to compute the loop amplitude from graphs omitting graphs with self-energy corrections on external lines.

\begin{figure}
\begin{center}
\includegraphics[scale=1.0]{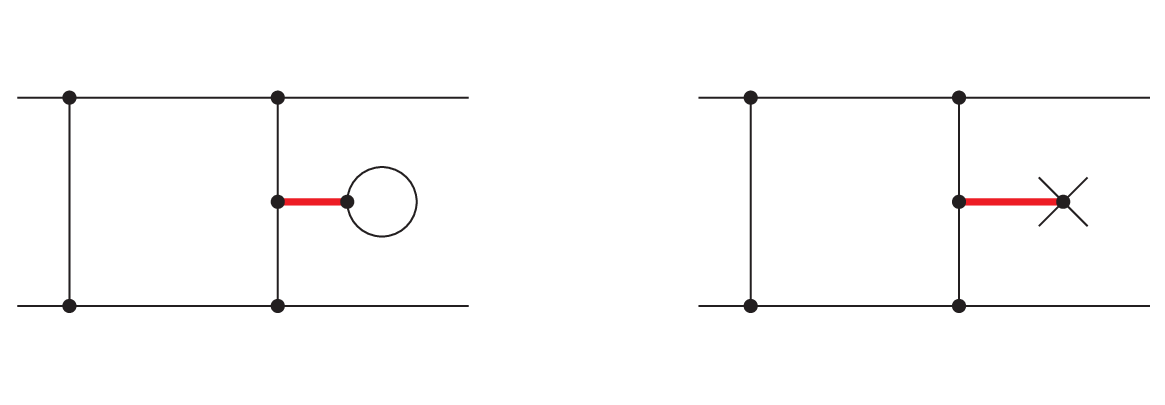}
\end{center}
\caption{
Two graphs with tadpoles.
These graph belongs to ${\mathcal U}_{2,4}$, but not to ${\mathcal U}_{2,4}^{\mathrm{no \; tadpoles}}$.
These graphs contain a zero-momentum propagator, indicated by a red line.
The counterterm in the right diagram corresponds to the renormalisation of the source $J$.
}
\label{fig_tadpole}
\end{figure}

\begin{figure}
\begin{center}
\includegraphics[scale=1.0]{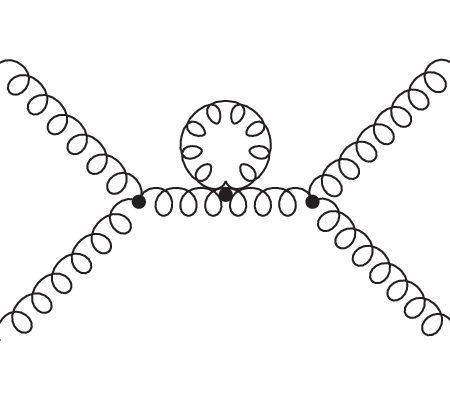}
\end{center}
\caption{
A snail graph in Yang-Mills theory. This is not a tadpole graph, since the internal edge with zero momentum is missing.
This graph belongs to ${\mathcal U}_{1,4}^{\mathrm{no \; tadpoles}}$.
}
\label{fig_snail}
\end{figure}

There is a second category of graphs, which we have to discuss more carefully.
These are graphs with tadpoles.
A graph with a tadpole contains a sub-graph without external edges, which is connected to the rest of the graph only by a single edge.
An example is shown in fig.~\ref{fig_tadpole}.
Graphs with tadpoles contain propagators with zero momentum. 
If the corresponding particle is massless, we would again be dividing by zero.
In quantum field theories, where the field $\phi$ has a vanishing vacuum expectation value, 
the one-point correlation function vanishes.
This implies that the sum of the two diagrams shown in fig.~\ref{fig_tadpole} gives zero,
where the counterterm in the second diagram corresponds to the renormalisation of the source $J$ \cite{Srednicki:2007qs}.
For this reason we included in eq.~(\ref{counterterms}) the renormalisation of the source $J$.
It is common practice to omit these diagrams in the calculation of loop amplitudes in theories where all fields have
a vanishing vacuum expectation value. They would simply add up to zero.
We denote the set of graphs without tadpoles by
\bq
 {\mathcal U}_{\loopnumber,n}^{\mathrm{no \; tadpoles}},
\eq 
and the set of graphs without self-energy corrections on external lines and without tadpoles by
\bq
 {\mathcal U}_{\loopnumber,n}^{\mathrm{amputated, \; no \; tadpoles}}.
\eq 
Let us remark that the graph within Yang-Mills theory shown in fig.~\ref{fig_snail} is not a tadpole graph 
and included in ${\mathcal U}_{\loopnumber,n}^{\mathrm{no \; tadpoles}}$.
Graphs like in fig.~\ref{fig_snail} are called snail graphs.
The graph in fig.~\ref{fig_snail} gives zero within dimensional regularisation.
In an analytic calculation these graphs are therefore often dropped.
However, this zero comes from a cancellation of an ultraviolet divergence with an infrafred divergence.
Within a numerical calculation we do not drop these graphs, since dropping them would result in mixing ultraviolet
divergences with infrared divergences.

\begin{figure}
\begin{center}
\includegraphics[scale=1.0]{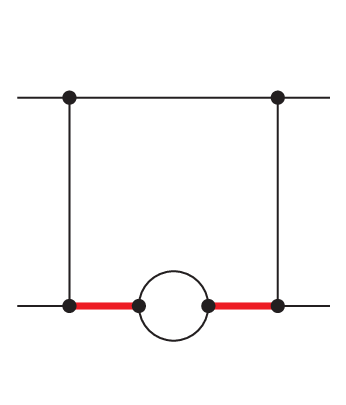}
\end{center}
\caption{
A graph with a self-energy insertion on an internal line.
This graph belongs to ${\mathcal U}_{2,4}$ and ${\mathcal U}_{2,4}^{\mathrm{amputated}}$, but not to
${\mathcal U}_{2,4}^{\mathrm{no \; self-energies}}$.
The graph contains a squared propagator, originating from the two red lines.
}
\label{fig_internal_self_energy}
\end{figure}

Finally, let us discuss a third category of graphs, which deserve special attention.
These are graphs with self-energy corrections on internal lines.
An example is shown in fig.~\ref{fig_internal_self_energy}.
Self-energy insertions on internal lines lead to higher powers of the propagators.
We denote by 
\bq
 {\mathcal U}_{\loopnumber,n}^{\mathrm{no \; self-energies}}
\eq
the subset of graphs without any self-energy insertions, external or internal
and by
\bq
 {\mathcal U}_{\loopnumber,n}^{\mathrm{no \; self-energies, \; no \; tadpoles}}
\eq
the subset of graphs without any external or internal self-energy insertions and no tadpoles.

Throughout this paper we assume that the ultraviolet counterterms have an integral representation, e.g. 
a counterterm for a propagator has an integral representation in the form of a two-point function,
a counterterm for a three-valent vertex has an integral representation in the form of a three-point function, etc.,
such that the sum of the integrand of the bare part and the integrand of the counterterm part is integrable in loop momentum space.
These integral representations for the counterterms may be constructed systematically \cite{Becker:2010ng,Becker:2012aq,Baumeister:2019rmh}.
This allows to treat graphs from ${\mathcal U}_{\loopnumber,n}$ without or with counterterms 
on the same footing collectively as graphs
with $n$ external particles and $\loopnumber$ loops.
If we want to refer explicitly to graphs without or with at least one counterterm we write
\bq
 {\mathcal U}_{\loopnumber,n}^{\mathrm{no \; CT}}
 & \mbox{or} &
 {\mathcal U}_{\loopnumber,n}^{\mathrm{CT}}
\eq
for the corresponding sets of graphs, respectively.
Obviously
\bq
 {\mathcal U}_{\loopnumber,n}^{\mathrm{no \; CT}}
 \cup
 {\mathcal U}_{\loopnumber,n}^{\mathrm{CT}}
 \; = \;
 {\mathcal U}_{\loopnumber,n},
 & &
 {\mathcal U}_{\loopnumber,n}^{\mathrm{no \; CT}}
 \cap
 {\mathcal U}_{\loopnumber,n}^{\mathrm{CT}}
 \; = \;
 \emptyset.
\eq
Sets of graphs with additional restrictions are defined analogously.
For example,
\bq
 {\mathcal U}_{\loopnumber,n}^{\mathrm{\mathrm{no \; CT}, \; no \; self-energies, \; no \; tadpoles}}
\eq
denotes the set of graphs without any ultraviolet counterterms, no (internal or external) self-energy insertions
and no tadpoles.

\section{Cutting and sewing}
\label{sect:cutting_sewing}

In this section we introduce two operations on graphs, cutting and sewing, which are inverse to each other.
For a graph $\Gamma \in {\mathcal U}_{\loopnumber,n}$ we denote by $E_\Gamma=\{e_1,...,e_N\}$ the set of internal edges.

Our first step is to introduce cut trees. Within graph theory there is a well established notion of spanning trees.
These two concepts are related, but not identical.
It is helpful to present both definitions, highlighting the similarities and the differences.

Let us first review the definition of a spanning tree \cite{Bogner:2010kv}.
A spanning tree for the graph $\Gamma$ is a sub-graph $T_{\mathrm{span}}$ of $\Gamma$, 
which contains all the vertices of $\Gamma$ and is a connected tree graph.
If $T_{\mathrm{span}}$ is a spanning tree for $\Gamma$, 
then it can be obtained from $\Gamma$ by deleting $\loopnumber$ internal edges, say $\{e_{\sigma_1},...,e_{\sigma_\loopnumber}\}$.
We denote by $\sigma=\{\sigma_1,...,\sigma_{\loopnumber}\}$ the set of indices of the deleted edges.
We denote by
\bq
 {\mathcal T}_\Gamma
\eq
the set of all spanning trees for the graph $\Gamma$ 
and by
\bq
 {\mathcal C}_\Gamma
\eq
the set of all sets of indices of the deleted edges.
There is a bijection between ${\mathcal T}_\Gamma$ and ${\mathcal C}_\Gamma$.
A spanning tree ${\mathcal T}_\Gamma$ for $\Gamma \in {\mathcal U}_{\loopnumber,n}$ has $n$ external lines.

Each $\sigma \in {\mathcal C}_\Gamma$ defines also a cut graph $T_{\mathrm{cut}}$, obtained by cutting
each of the $\loopnumber$ internal edges $e_{\sigma_j}$ into two half-edges. 
The $2\loopnumber$ half-edges become external lines of $T_{\mathrm{cut}}$.
The graph $T_{\mathrm{cut}}$ is a tree graph with $n+2\loopnumber$ external lines.
The difference between a spanning tree and a cut tree 
\begin{figure}
\begin{center}
\includegraphics[scale=1.0]{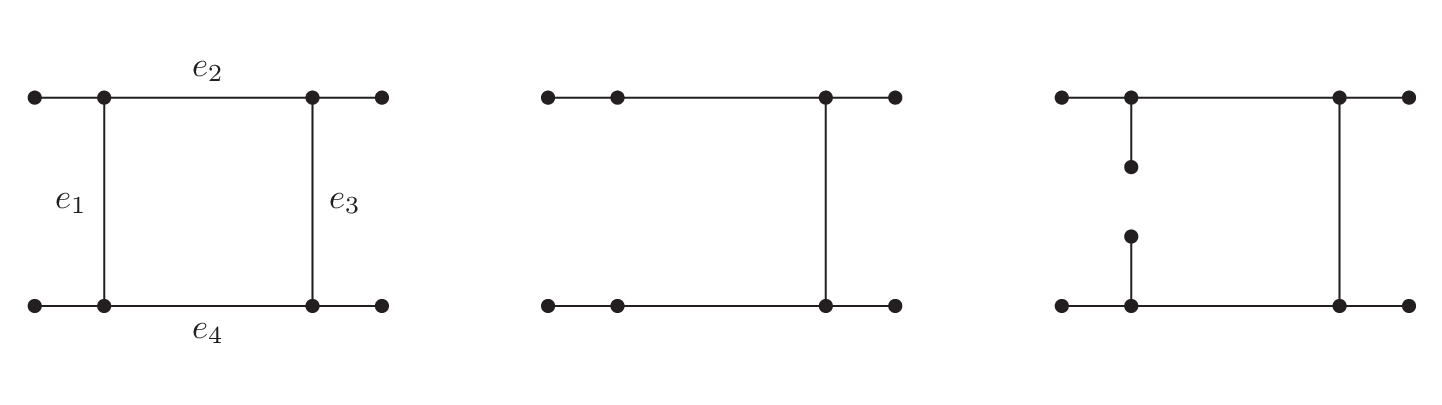}
\end{center}
\caption{
The left picture shows a graph $\Gamma$ from ${\mathcal U}_{1,4}$.
The middle picture shows a spanning tree for this graph, obtained by deleting the internal edge $e_1$.
The spanning tree has four external lines.
The right picture shows the corresponding cut tree, obtained by cutting the internal edge $e_1$.
The cut tree has six external lines.
In these figures we included a vertex at the end of each external line. 
Usually the vertices at the end of external lines are not drawn.
}
\label{fig_spanning_tree_cut_tree}
\end{figure}
is illustrated in fig.~\ref{fig_spanning_tree_cut_tree}.

Further, we denote by $U_\Gamma$ the first graph polynomial of the graph $\Gamma$.
In order to obtain $U_\Gamma$, we associate to each internal edge $e_j$ a variable $x_j$.
The graph polynomial $U_\Gamma$ is a homogeneous polynomial of degree $\loopnumber$ in the variables $x_j$,
obtained as a sum over monomials with coefficients $+1$.
Each monomial consists of a product of $x_j$'s, such that when the corresponding edges are deleted (or cut),
we obtain a connected tree graph.
The sum is over all those possibilities.
Put into a formula, we have
\bq
 U_\Gamma
 & = & 
 \sum\limits_{T\in {\mathcal T}_\Gamma} \;
     \prod\limits_{e_i\notin T} x_i.
\eq
The number of spanning trees for a graph $\Gamma$ is given by
\bq
 \left| {\mathcal T}_\Gamma \right|
 \;\; = \;\;
 \left| {\mathcal C}_\Gamma \right|
 \;\; = \;\;
 \left. U_\Gamma\right|_{x_1=...=x_N=1}.
\eq 
At the level of graphs, cutting an internal edge $e_{\sigma_j}$ of a graph $\Gamma \in {\mathcal U}_{\loopnumber,n}$
yields a graph with $(\loopnumber-1)$ loops and $n+2$ external legs.
Repeating this $\loopnumber$ times, gives a tree graph with $n+2\loopnumber$ external legs.

At the level of graphs, sewing is the inverse operation of cutting.
Consider a graph $\Gamma \in {\mathcal U}_{\loopnumber-1,n+2}$ with $(\loopnumber-1)$ loops and 
$(n+2)$ external legs.
We label the external legs by
\bq
 (p_1,p_2,...,p_n,k_1,\bar{k}_1).
\eq
Sewing the external edges labelled by $k_1$ and $\bar{k}_1$ means connecting the two external edges to form
a new internal edge.
The resulting graph has then $\loopnumber$ loops and $n$ external lines.
Starting with a graph $\Gamma \in {\mathcal U}_{0,n+2\loopnumber}$ with external edges
labelled by 
\bq
 (p_1,p_2,...,p_n,k_1,...,k_\loopnumber,\bar{k}_1,...,\bar{k}_\loopnumber)
\eq
we may repeat the sewing procedure $\loopnumber$ times and sew the external edge labelled by $k_j$
with the external labelled by $\bar{k}_j$ for $j=1,...,\loopnumber$.
The resulting graph has then $\loopnumber$ loops and $n$ external edges.

For the moment we discussed the operations of cutting and sewing purely at the level of graphs.
Of course, we have in mind that each Feynman graph represents a mathematical expression.
The translation from graphs to mathematical expressions is given by the Feynman rules.
At the level of mathematical expressions, the cutting operation corresponds to taking the residue when
the cut propagator goes on-shell.
On the other hand, the sewing operation only makes sense in the case $\bar{k}_j=-k_j$.
The sewing operation corresponds to the forward limit.
We remind the reader that by convention we take all momenta outgoing.
The outgoing momentum $\bar{k}_j$ corresponds to the incoming momentum $-\bar{k}_j$.
In the case $\bar{k}_j=-k_j$, the incoming momentum equals $-\bar{k}_j=k_j$.
Thus, the particle with outgoing momentum $\bar{k}_j$ has the incoming momentum $k_j$, which coincides
with the outgoing momentum $k_j$ of the other particle.
This is the forward limit.

In theories with spin the sewing operation implies also a sum over the physical and unphysical polarisations 
of the particles corresponding to the sewed edges.
For example, 
for spin $1/2$-fermions we have
\bq
\label{pol_sum_fermions}
\sum\limits_{\lambda} u^\lambda\left(k\right) \bar{u}^\lambda\left(k\right) = k\!\!\!/ + m,
 & &
\sum\limits_{\lambda} v^\lambda\left(k\right) \bar{v}^\lambda\left(k\right) = k\!\!\!/ - m.
\eq
For massless gauge bosons of spin $1$ the sum over the physical polarisations gives
\bq
 \sum\limits_{\lambda} \left( \eps_\mu^\lambda\left(k,n\right) \right)^\ast \eps_\nu^\lambda\left(k,n\right)
 & = &
 - g_{\mu\nu} + \frac{k_\mu n_\nu + n_\mu k_\nu}{k \cdot  n},
\eq
where $n$ is a light-like reference vector.
We would like the sewing operation to reproduce the numerator of the propagator.
For gauge bosons this numerator is gauge-dependent.
In Feynman gauge the numerator is given by $(-g_{\mu\nu})$.
We may write
\bq
\label{pol_sum_boson}
 - g_{\mu\nu} 
 & = &
 \sum\limits_{\lambda} \left( \eps_\mu^\lambda\left(k,n\right) \right)^\ast \eps_\nu^\lambda\left(k,n\right)
 - \frac{k_\mu n_\nu + n_\mu k_\nu}{k \cdot  n},
\eq
expressing the Feynman gauge numerator as a sum over physical and unphysical polarisations.
The latter are proportional to $k_\mu$ and $n_\mu$.
In the same spirit we include in gauge theories diagrams, where we sew together a ghost particle with its 
corresponding anti-ghost.

In addition, we define the sewing operation to include a minus sign 
\bq
 \left(-1\right)
\eq
for each sewing of a fermion line.
Please note that this minus sign is included independently if the sewing operation closes a fermion loop or not.
Ghost lines are treated as fermion lines.

\section{Loop-tree duality}
\label{sect:loop_amplitudes}

Let us consider loop amplitudes in a quantum field theory, where all fields have vanishing vacuum expectation values.
In this case we may ignore tadpole contributions and our relevant set of diagrams is
\bq
 {\mathcal U}_{\loopnumber,n}^{\mathrm{loop}}
 & = &
 {\mathcal U}_{\loopnumber,n}^{\mathrm{amputated, \; no \; tadpoles}}.
\eq 
Self-energy insertions on internal lines contribute to the loop amplitude and are included.
Let $\Gamma \in {\mathcal U}_{\loopnumber,n}^{\mathrm{loop}}$ 
and $E_\Gamma=\{e_1,...,e_N\}$ the set of internal edges.
For each internal edge we set
\bq
 D_j & = & k_j^2 - m_j^2 + i \delta, 
 \;\;\;\;\;\;
 e_j \in E_\Gamma,
\eq
where $\delta>0$ is an infinitesimal small quantity.
For gauge theories we choose Feynman gauge in order to avoid additional spurious poles.
Without loss of generality we may assume that we labelled the internal edges in such a way that
$k_1$, ..., $k_\loopnumber$ is a set of $\loopnumber$ independent loop momenta for the Feynman diagram $\Gamma$.
To each graph $\Gamma$ we associate the integrand
\bq
 f\left(\Gamma\right)
 & = &
 \frac{P_\Gamma}{\prod\limits_{e_j \in E_\Gamma} D_j}.
\eq
$P_\Gamma$ is a polynomial in the independent loop momenta $k_1$, ..., $k_\loopnumber$
and the external momenta $p_1$, ..., $p_n$.
The $\loopnumber$-loop amplitude with $n$ external legs is the sum over all contributing Feynman diagrams
\bq
\label{loop_amplitude_Feynman_diagrams}
 {\mathcal A}_{\loopnumber,n}\left(p_1,...,p_n\right)
 & = &
 \sum\limits_{\Gamma \in {\mathcal U}_{\loopnumber,n}^{\mathrm{loop}}}
 \frac{\left(-1\right)^{l_{\mathrm{cfl}}(\Gamma)}}{\left|\mathrm{Aut}\left(\Gamma\right)\right|}
 \int \left( \prod\limits_{j=1}^\loopnumber \frac{d^Dk_j}{\left(2\pi\right)^D} \right)
 f\left(\Gamma\right),
\eq
where $1/|\mathrm{Aut}(\Gamma)|$ denotes the symmetry factor of the diagram $\Gamma$
and $l_{\mathrm{cfl}}(\Gamma)$ the number of closed fermion loops in $\Gamma$.
Within dimensional regularisation, the $\loopnumber$-loop integral is translation invariant.
For each Feynman diagram, we have quite some freedom in choosing the loop momenta $k_1$, ..., $k_\loopnumber$.
We may shift the loop momenta by $(D \cdot \loopnumber)$ translations.
In addition, we may perform a ${\mathrm SL}(\loopnumber,{\mathbb R})$-transformation
(or a  ${\mathrm SL}(\loopnumber,{\mathbb Z})$-transformation, if we want to preserve the property, that each internal
momentum is a linear combination of the external momenta and the independent loop momenta with integer coefficients)
on the $\loopnumber$ independent loop momenta.

Our aim is to define an integrand of the loop amplitude 
(i.e. exchange the summation and the integration in eq.~(\ref{loop_amplitude_Feynman_diagrams})), 
such that the integrand {\sw becomes a tree-amplitude-like object, where we expect ``nice'' factorisation properties in all infrared limits.}
The integrand of a $\loopnumber$-loop amplitude should be a rational function in
the loop momenta $k_1$, ..., $k_\loopnumber$
and the external momenta $p_1$, ..., $p_n$.

Defining ``some'' integrand is easy: 
For each Feynman diagram we may choose a set of independent loop momenta, relabel them $k_1$, ..., $k_\loopnumber$,
and add up the contributions from the individual Feynman diagrams including the symmetry factors and the minus signs for each closed fermion loop.
This will give a rational function in
the loop momenta $k_1$, ..., $k_\loopnumber$ (defined as independent loop momenta for all diagrams)
and the external momenta $p_1$, ..., $p_n$.
However, in general this rational function will not have ``nice'' factorisation properties in all infrared limits.

We do not know if this is actually possible at the level of $D$-dimensional off-shell loop momenta $k_1$, ..., $k_\loopnumber$.
In order to proceed, we chop each Feynman diagram into several pieces.
Using loop-tree duality we rewrite each Feynman integral as a sum over $\loopnumber$-fold residues from the energy integrations.
We then use translation invariance in the remaining spatial integrations for each piece individually.
Finally, we re-assemble the pieces, which gives us the regularised $\loopnumber$-fold forward limit of a {\sw tree-amplitude-like object}.

Within this approach we have to address a few technical complications:
\begin{enumerate}
\item The $\loopnumber$-fold forward limit of a tree amplitude is in general a singular limit.
\item {\sw Symmetry factors, minus signs for closed fermion loops and combinatorial factors}.
\item Loop diagrams with higher powers of the propagators.
\end{enumerate}
We will address these challenges in the next sections.

Let us start with a review of loop-tree duality \cite{Catani:2008xa,Runkel:2019yrs}.
For a function $f$ depending on a $D$-dimensional momentum variable $k=(E,\vec{k})$,
where the vector $\vec{k}$ is $(D-1)$-dimensional,
we either write $f(k)$ or $f(E,\vec{k})$.
Within the loop-tree duality approach we perform the energy integration with the help of Cauchy's residue theorem.
This leaves the integration over the spatial components $\vec{k}$.
As a short hand notation we write for cut graphs
\bq
\label{def_forward_backward}
 \int\limits_{+/-} \frac{d^{D-1}k}{\left(2\pi\right)^{D-1} 
} 
 \;
 f\left(k\right)
 & = &
 \int \frac{d^{D-1}k}{\left(2\pi\right)^{D-1} 
} 
 \left[ 
 f\left(\sqrt{\vec{k}^2+m^2},\vec{k}\right)
 +
 f\left(-\sqrt{\vec{k}^2+m^2},\vec{k}\right)
 \right]
\eq
for the integral over the forward and the backward hyperboloid.
Let $\Gamma \in {\mathcal U}_{\loopnumber,n}$ and let $\sigma \in {\mathcal C}_\Gamma$ 
be a set of indices defining a spanning tree.
For each cut edge we choose an orientation and we may take the $\loopnumber$ independent loop momenta to be the loop momenta flowing
through the edges $e_{\sigma_1}, ..., e_{\sigma_\loopnumber}$ with the chosen orientation.
Let
\bq
 E_\sigma^{(\alpha)} & = & \left(E_{\sigma_1}^{(\alpha)},...,E_{\sigma_l}^{(\alpha)}\right)
\eq
be a solution to
\bq
 D_{\sigma_1}
 \;\; = \;\;
 ...
 \;\; = \;\;
 D_{\sigma_\loopnumber}
 \;\; = \;\;
 0.
\eq
In total there are $2^\loopnumber$ solutions {\sw $E_\sigma^{(\alpha)}$, indexed by $\alpha =(\alpha_1,...,\alpha_l) \in \{1,-1\}^l$ and }
given by
\bq
 \left( \pm \sqrt{ \vec{k}_{\sigma_1}^2 + m_{\sigma_1}^2 - i \delta}, ..., \pm \sqrt{ \vec{k}_{\sigma_\loopnumber}^2 + m_{\sigma_\loopnumber}^2 - i \delta} \right).
\eq
Let us denote by $n_\sigma^{(\alpha)}$ the number of times the negative root $-\sqrt{...}$ occurs in
$E_\sigma^{(\alpha)}$.
We define the local residue \cite{Griffiths:book} at $E_\sigma^{(\alpha)}$ by
\bq
\label{local_residue}
 \mathrm{res}\left(f,E_\sigma^{(\alpha)}\right)
 & = &
 \frac{1}{\left(2\pi i\right)^l}
 \oint\limits_{\gamma_\eps} f dE_{\sigma_1} \wedge ... \wedge dE_{\sigma_\loopnumber}.
\eq
The integration in eq.~(\ref{local_residue}) is around a small $\loopnumber$-torus
\bq
 \gamma_\eps & = &
 \left\{
   \left( E_{\sigma_1}, ..., E_{\sigma_\loopnumber} \right) \in {\mathbb C}^\loopnumber | \left| D_{\sigma_i} \right| = \eps
 \right\},
\eq
encircling $E_\sigma^{(\alpha)}$ with orientation
\bq
 d \arg D_{\sigma_1} \wedge d \arg D_{\sigma_2} \wedge ... \wedge d \arg D_{\sigma_\loopnumber} \ge 0.
\eq
Loop-tree duality allows to re-write a $\loopnumber$-loop integral as
\bq
\label{general_loop_tree_duality}
 \int \left( \prod\limits_{j=1}^\loopnumber \frac{d^Dk_j}{\left(2\pi\right)^D} \right)
 f\left(\Gamma\right)
 & = &
{\sw
 \left(-i\right)^l
 \sum\limits_{\sigma \in {\mathcal C}_\Gamma}
 \int\limits_{+/-} \left( \prod\limits_{j=1}^\loopnumber \frac{d^{D-1}k_{\sigma_j}}{\left(2\pi\right)^{D-1}} \right)
 S_{\sigma \alpha}
 \left(-1\right)^{n_\sigma^{(\alpha)}}
 \mathrm{res}\left(f,E_\sigma^{(\alpha)}\right).
}
\eq
The sum is over all cut trees of $\Gamma$. 
If all propagators occur to power one, eq.~(\ref{general_loop_tree_duality}) simplifies to
\bq
\label{loop_tree_duality_single_poles}
\lefteqn{
 \int \left( \prod\limits_{j=1}^\loopnumber \frac{d^Dk_j}{\left(2\pi\right)^D} \right)
 \frac{P_\Gamma}{\prod\limits_{e_j \in E_\Gamma} \left( k_j^2 - m_j^2 + i \delta \right)}
 = 
 } & &
 \nonumber \\
 & &
{\sw
 \left(-i\right)^\loopnumber
 \sum\limits_{\sigma \in {\mathcal C}_\Gamma}
 \int\limits_{+/-} \left( \prod\limits_{j=1}^\loopnumber \frac{d^{D-1}k_{\sigma_j}}{\left(2\pi\right)^{D-1} 2 \sqrt{\vec{k}_{\sigma_j}^2 + m_{\sigma_j}^2 }} \right)
 S_{\sigma \alpha}
 \frac{P_\Gamma}{\prod\limits_{j \notin \sigma } \left( k_j^2 - m_j^2 + i s_j\left(\sigma\right) \delta \right)}.
}
\eq
The propagators corresponding to the edges $e_{\sigma_1}$, ..., $e_{\sigma_\loopnumber}$ are cut,
the remaining propagators have a modified (``dual'') $i\delta$-prescription.
The quantity $s_j(\sigma)$ is defined by \cite{Runkel:2019yrs}
\bq
\label{dual_delta_I_prescription}
 s_j\left(\sigma\right)
 & = &
 \frac{E_j}{E_\parallel}
\eq
and $E_\parallel$ is defined as follows:
The set $\sigma = \{\sigma_1,...,\sigma_\loopnumber\} \in {\mathcal C}_\Gamma$ defines 
a tree $T_{\mathrm{cut}}$ obtained from the graph $\Gamma$ by cutting the internal edges 
$C_\sigma = \{e_{\sigma_1},...,e_{\sigma_\loopnumber}\}$.
Cutting in addition the edge $e_j \in E_\Gamma \backslash C_\sigma$ will give a two-forest $(T_1,T_2)$.
We orient the external momenta of $T_1$ such that all momenta are outgoing.
Let $\pi$ be the set of indices corresponding to the external edges of $T_1$ which come from cutting the edges
$C_\sigma$ of the graph $\Gamma$.
The set $\pi$ may contain an index twice, this is the case if both half-edges of a cut edge belong to $T_1$.
Then define $E_\parallel$ by
\bq
 \frac{1}{E_\parallel}
 & = &
 \sum\limits_{a \in \{j\} \cup \pi} \frac{1}{E_a}.
\eq
Although we singled out the tree $T_1$ from the two-forest $(T_1,T_2)$ it is easily checked that the definition
of $s_j(\sigma)$ is invariant under the exchange $T_1 \leftrightarrow T_2$.

{\sw
$S_{\sigma \alpha}$ is a combinatorial factor.
The origin of the combinatorial factor is as follows \cite{Capatti:2019ypt}:
We would like to write the left-hand side of eq.~(\ref{general_loop_tree_duality}) as a sum of local residues
indexed by $\sigma$ and $\alpha$.
Such a representation as a sum of local residues is not unique. This is due to the fact that the sum of all residues in any subloop equals
zero.
We obtain a well-defined representation as a sum of local residues by specifying
a set of integration variables by $\tilde{\sigma} \in {\mathcal C}_\Gamma$,
an order in which the integrations are performed by $\tilde{\pi} \in S_l$ 
and a ordered set of winding numbers $\tilde{\alpha}=(\Gamma_1,...,\Gamma_l)$.
We assume that the integration over $k_{\tilde{\sigma}_{\tilde{\pi_1}}}$ is performed first, followed
by the integration over $k_{\tilde{\sigma}_{\tilde{\pi_2}}}$, etc..
A positive winding number implies that the corresponding integration contour is closed above, 
for a negative winding number it is closed below.
In order to keep the indexing to a minimum we introduce the ordered set
$\tilde{k}=(\tilde{k}_1,...,\tilde{k}_l)=(k_{\tilde{\sigma}_{\tilde{\pi_1}}},...,k_{\tilde{\sigma}_{\tilde{\pi_l}}})$.
For a cut specified by $\sigma \in {\mathcal C}_\Gamma$ we denote by $\pi \in S_l$ the order in which the cuts
are taken, e.g. the cut of the edge $e_{\sigma_{\pi_1}}$ is taken in the first integration, followed
by the the cut of the edge $e_{\sigma_{\pi_2}}$, etc..
Again, in order to keep the indexing to a minimum we introduce the ordered set
$\hat{k}=(\hat{k}_1,...,\hat{k}_l) = (k_{\sigma_{\pi_1}}, ..., k_{\sigma_{\pi_l}})$.
As before, we denote by $\alpha=(\alpha_1,...,\alpha_l)$ the signs of the energies for the cut under consideration.
$\alpha_j=1$ means that we consider the residue with positive energy with respect to the chosen
orientation of the edge $e_{\sigma_{\pi_j}}$.
$\hat{k}$ and $\tilde{k}$ are both bases of independent loop momenta, hence they are related by
\bq
 \hat{k}_i & = & \sum\limits_{j=1}^l \Sigma_{ij} \tilde{k}_j + q_i,
\eq
with $\Sigma_{ij} \in \{-1,0,1\}$ and $q_i$ depending only on the external momenta.
This defines the $l \times l$-signature matrix $\Sigma$.
We denote by $\Sigma^{(j)}$ the $j \times j$-matrix obtained from $\Sigma$ by deleting the rows and columns $(j+1), ..., l$.
In order to compute the residues we may temporarily assume that the imaginary parts of all internal masses are large and strongly ordered.
The final result will not depend on this assumption.
After performing the contour integrations we may remove this assumption
and analytically continue to any desired (complex) kinematics.
With these specifications one obtains
\bq
\label{corrected_result}
 \frac{1}{\left(2\pi i\right)^l}
 \int f\left(\Gamma\right) dE_1 \wedge ... \wedge dE_l
 & = &
 \sum\limits_{\sigma \in {\mathcal C}_\Gamma}
 \sum\limits_{\pi \in S_l}
 \sum\limits_{\alpha \in \{1,-1\}^l}
 C^{\tilde{\sigma} \tilde{\pi} \tilde{\alpha}}_{\sigma \pi \alpha}
 \;
 \mathrm{res}\left(f,E_\sigma^{(\alpha)}\right),
\eq
where
$C^{\tilde{\sigma} \tilde{\pi} \tilde{\alpha}}_{\sigma \pi \alpha}$ is given by
\bq
 C^{\tilde{\sigma} \tilde{\pi} \tilde{\alpha}}_{\sigma \pi \alpha}
 & = &
 \prod\limits_{i=1}^l \Delta^{(i)}.
\eq
$\Delta^{(i)}$ is zero if $\det \Sigma^{(i)} = 0$. Otherwise we let $\Pi^{(i)}$ be the inverse matrix of $\Sigma^{(i)}$.
The quantity $\Delta^{(i)}$ is then given by
\bq
 \Delta^{(i)}
 & = &
 \Gamma_i \Pi^{(i)}_{ii}
 \;
 \theta\left( \Gamma_i \mathrm{Im}\left( \sum\limits_{j=1}^i \Pi^{(i)}_{ij} \alpha_j m_{\sigma_{\pi_j}} \right) \right).
\eq
The quantities $\Delta^{(i)}$ are computed with a chosen strong ordering of the imaginary parts of the internal masses.
The quantity $C^{\tilde{\sigma} \tilde{\pi} \tilde{\alpha}}_{\sigma \pi \alpha}$ is independent of this choice.

This gives a well-defined representation in terms of a sum of local residues, but this representation depends
on our choice of $\tilde{\sigma}$, $\tilde{\pi}$ and $\tilde{\alpha}$.
One may now sum over $\pi$ and average over $\tilde{\sigma}$, $\tilde{\alpha}$, $\tilde{\pi}$ 
in a suitable way.
We do this as follows:
We group the internal propagators $D_j$ into chains \cite{Kinoshita:1962ur}.
Two propagators belong to the same chain, if their momenta differ only by a linear combination 
of the external momenta. We denote by $n^{\mathrm{chain}}(j)$ the number of propagators in the chain of $D_j$.
We set
\bq
 N^{\mathrm{chain}}\left(\sigma\right)
 & = &
 \prod\limits_{j=1}^l n^{\mathrm{chain}}\left(\sigma_j\right).
\eq
To each graph $\Gamma$ we associate a new graph $\Gamma^{\mathrm{chain}}$ called the chain graph by deleting all external lines
and by choosing one propagator for each chain as a representative.
We denote by $|{\mathcal C}_{\Gamma^{\mathrm{chain}}}|$ the number of spanning trees of the chain graph.
We then perform a weighted average, where each term is weighted by $1/N^{\mathrm{chain}}(\sigma)$.
We obtain
\bq
\label{result_with_combinatorial_factor}
 \frac{1}{\left(2\pi\right)^l}
 \int f dE_1 \wedge ... \wedge dE_l
 & = &
 \left(-i\right)^l
 \sum\limits_{\sigma \in {\mathcal C}_\Gamma}
 \sum\limits_{\alpha=1}^{2^l}
 S_{\sigma \alpha}
 \left(-1\right)^{n_\sigma^{(\alpha)}}
 \mathrm{res}\left(f,E_\sigma^{(\alpha)}\right),
\eq
with
\bq
 S_{\sigma \alpha}
 & = &
 \frac{\left(-1\right)^{l+n_\sigma^{(\alpha)}}}{2^l l! \left|{\mathcal C}_{\Gamma^{\mathrm{chain}}}\right|}
 \sum\limits_{\pi \in S_l}
 \sum\limits_{\tilde{\sigma} \in {\mathcal C}_\Gamma}
 \sum\limits_{\tilde{\pi} \in S_l}
 \sum\limits_{\tilde{\alpha} \in \{1,-1\}^l}
 \frac{C^{\tilde{\sigma} \tilde{\pi} \tilde{\alpha}}_{\sigma \pi \alpha}}{N^{\mathrm{chain}}\left(\sigma\right)}.
 \nonumber
\eq
This defines the $S_{\sigma \alpha}$ and ensures that the combinatorial factor $S_{\sigma \alpha}$
only depends on the underlying chain graph $\Gamma^{\mathrm{chain}}$.

Let us stress that any specific choice of $\tilde{\sigma}$, $\tilde{\pi}$, $\tilde{\alpha}$
in eq.~(\ref{corrected_result}) gives always the same function in terms of a fixed basis of the remaining 
spatial loop momenta, identical to the right-hand side of eq.~(\ref{result_with_combinatorial_factor}).
The averaging procedure is useful, when a sum over all diagrams is considered.
It allows to combine diagrams into off-shell currents.
}

On the right-hand side of eq.~(\ref{general_loop_tree_duality}) and eq.~(\ref{loop_tree_duality_single_poles})
we may relabel the loop integration momenta $( k_{\sigma_1}, k_{\sigma_2}, ..., k_{\sigma_\loopnumber} )$ to $(k_1,k_2,...,k_\loopnumber)$.
There are $\loopnumber!$ possibilities to do that and we may average over all of them.
This introduces a sum over all permutations from the symmetric group $S_\loopnumber$ together with a prefactor $1/\loopnumber!$.
{\sw
Please note that the relabeling of the loop momentum destroys the dual cancellations \cite{Buchta:2014dfa,Aguilera-Verdugo:2019kbz,Capatti:2019ypt} 
on H-surfaces among the different terms in the sum
on the right-hand side of eq.~(\ref{general_loop_tree_duality}) and eq.~(\ref{loop_tree_duality_single_poles}).
}

In theories with spin we replace the numerators of the cut propagators by polarisation sums.
This is straightforward for spin $1/2$-fermions.
The corresponding formulae are given in eq.~(\ref{pol_sum_fermions}).
The propagator of a (massless) spin $1$-gauge boson is gauge-dependent.
In order to avoid additional poles or additional higher poles it is advantageous to work in Feynman gauge,
where the numerator of the gauge boson propagators is given by $(-g_{\mu\nu})$.
It is not possible, to express $(-g_{\mu\nu})$ as a sum over physical polarisations, but we may express it as a sum
over physical and unphysical polarisations as in eq.~(\ref{pol_sum_boson}).
In addition there are in gauge theories diagrams, which are obtained from cutting ghost lines.

\section{The regularised forward limit}
\label{sect:forward_limit}

In this section we define the regularised $\loopnumber$-fold forward limit of a tree amplitude {\sw and tree-amplitude-like objects}.
Let
\bq
 \left( p_1, ..., p_n, k_1, ..., k_\loopnumber, \bar{k}_1, ..., \bar{k}_\loopnumber \right)
\eq
be a set of $(n+2\loopnumber)$ external on-shell momenta, satisfying momentum conservation.
We denote the masses of the external particles by
\bq
 \left( m_j^{\mathrm{ext}} \right)^2
 \; = \; p_j^2,
 & &
 m_j^2 \; = \; k_j^2 \; = \; \bar{k}_j^2.
\eq
Let
\bq
 {\mathcal A}_{0,n+2\loopnumber}\left( p_1, ..., p_n, k_1, ..., k_\loopnumber, \bar{k}_1, ..., \bar{k}_\loopnumber \right)
 & = &
 \sum\limits_{\Gamma \in {\mathcal U}_{0,n+2\loopnumber}}
 f\left(\Gamma\right),
\eq
be the corresponding tree amplitude.
We are interested in the $\loopnumber$-fold forward limit
\bq
 \lim\limits_{\bar{k}_1 \rightarrow - k_1} ... \lim\limits_{\bar{k}_\loopnumber \rightarrow - k_\loopnumber}
 {\mathcal A}_{0,n+2\loopnumber}
\eq
This limit is singular for two reasons:
First, there are diagrams in ${\mathcal U}_{0,n+2\loopnumber}$, which give in the forward limit an on-shell propagator.
\begin{figure}
\begin{center}
\includegraphics[scale=1.0]{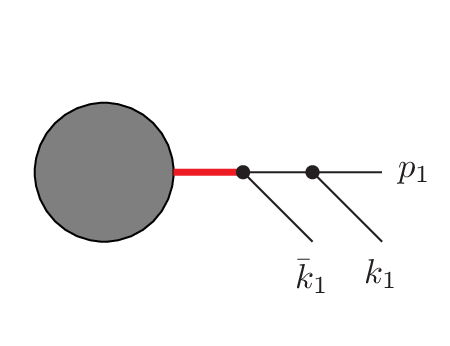}
\end{center}
\caption{
A diagram with a singular forward limit. In the limit $\bar{k}_1 \rightarrow -k_1$ the red propagator goes on-shell.
The blob represents the rest of the diagram.
Sewing $k_1$ with $\bar{k}_1$ gives a self-energy insertion on an external line.
}
\label{fig_forward_limit_on_shell}
\end{figure}
This case is further be divided into two sub-cases.
The first sub-case is characterised as follows: Let $\alpha$ be a subset of $\{1,...,\loopnumber\}$.
Diagrams, which have an internal edge with momentum
\bq
\label{condition_forward_on_shell}
 p_j + \sum\limits_{a \in \alpha} \left( k_a + \bar{k}_a \right),
 \;\;\;\;\;\;\;\;\;
 j \in \{1,...,n\}
\eq
and mass $m_j^{\mathrm{ext}}$
are singular in the forward limit.
An example for the first sub-case is shown in fig.~\ref{fig_forward_limit_on_shell}.
Sewing $k_a$ with $\bar{k}_a$ will give a diagram with a self-energy insertion on an external line.

In the second sub-case we replace $p_j$ in eq.~(\ref{condition_forward_on_shell}) by $k_b$ or $\bar{k}_b$ with $b \in \{1,...,\loopnumber\}\backslash \alpha$.
The second sub-case is characterised by diagrams,
\begin{figure}
\begin{center}
\includegraphics[scale=1.0]{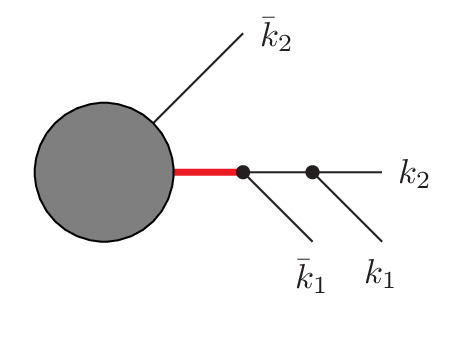}
\end{center}
\caption{
A diagram with a singular forward limit. In the limit $\bar{k}_1 \rightarrow -k_1$ the red propagator goes on-shell.
The blob represents the rest of the diagram.
Sewing $k_1$ with $\bar{k}_1$ and $k_2$ with $\bar{k}_2$ gives a self-energy insertion on an internal line and corresponds
to a graph with higher powers of a propagator.
}
\label{fig_forward_limit_higher_power}
\end{figure}
which have an internal edge with momentum
\bq
\label{condition_forward_higher_power}
 k_b + \sum\limits_{a \in \alpha} \left( k_a + \bar{k}_a \right)
 & \mbox{or} &
 \bar{k}_b + \sum\limits_{a \in \alpha} \left( k_a + \bar{k}_a \right),
 \;\;\;\;\;\;\;\;\;
 b \in \{1,...,\loopnumber\}\backslash \alpha,
\eq
and mass $m_b$.
An example is shown in fig.~\ref{fig_forward_limit_higher_power}.
Sewing $k_a$ with $\bar{k}_a$ for $a \in \{1,...,\loopnumber\}$ 
will give a diagram with a self-energy insertion on an internal line
and corresponds to a graph with higher powers of a propagator.

Our second principal case are diagrams in ${\mathcal U}_{0,n+2\loopnumber}$, which give in the forward limit a propagator with zero momentum.
\begin{figure}
\begin{center}
\includegraphics[scale=1.0]{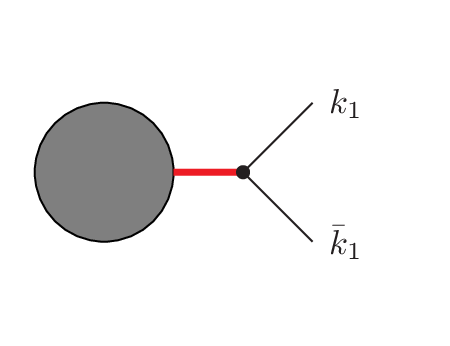}
\end{center}
\caption{
A diagram with a singular forward limit. In the limit $\bar{k}_1 \rightarrow -k_1$ the red propagator becomes a zero-momentum propagator.
The blob represents the rest of the diagram.
Sewing $k_1$ with $\bar{k}_1$ gives a tadpole.
}
\label{fig_forward_limit_zero_momentum}
\end{figure}
These are diagrams, which have an internal edge with momentum
\bq
\label{condition_forward_zero_momentum}
 \sum\limits_{a \in \alpha} \left( k_a + \bar{k}_a \right).
\eq
An example is shown in fig.~\ref{fig_forward_limit_zero_momentum}.
Sewing $k_a$ with $\bar{k}_a$ gives a tadpole.
If the internal edge with the momentum given in eq.~(\ref{condition_forward_zero_momentum})
has zero mass, we have again a singular forward limit.
(In the case of a non-zero mass, but a vanishing vacuum expectation value of the corresponding field, we may as well
ignore this contribution: Sewing gives a tadpole, which cancels with the counterterm from the renormalisation of the source.)

Let us therefore define the set of diagrams
\bq
 {\mathcal U}_{0,n+2\loopnumber}^{\mathrm{non-singular}}
\eq
as the subset of ${\mathcal U}_{0,n+2\loopnumber}$ without the singular graphs.
More concretely, these are all graphs from ${\mathcal U}_{0,n+2\loopnumber}$, except the ones which contain
a propagator with momentum of the form as in eq.~(\ref{condition_forward_on_shell}), (\ref{condition_forward_higher_power}) or (\ref{condition_forward_zero_momentum}).
We define the regularised $\loopnumber$-forward limit as
\bq
\label{def_forward_limit}
 {\mathcal R}_f
 {\mathcal A}_{0,n+2\loopnumber}
 & = &
 \lim\limits_{\bar{k}_1 \rightarrow - k_1} ... \lim\limits_{\bar{k}_\loopnumber \rightarrow - k_\loopnumber}
 \sum\limits_{\Gamma \in {\mathcal U}_{0,n+2\loopnumber}^{\mathrm{non-singular}}}
 f\left(\Gamma\right).
\eq
This defines the regularised $\loopnumber$-fold forward limit in terms of Feynman diagrams.

In gauge theories we allow for the external particles labelled by
\bq
 \left( k_1, ..., k_\loopnumber, \bar{k}_1, ..., \bar{k}_\loopnumber \right)
\eq
to have physical and unphysical polarisations in accordance with eq.~(\ref{pol_sum_boson}).
Furthermore, these particles are allowed to be ghosts or anti-ghosts.

{\sw
In the following sections we will work with a tree-amplitude-like object, obtained from a tree amplitude by the replacement
\bq
 f\left(\Gamma\right) & \rightarrow & S_{\sigma \alpha} \; f\left(\Gamma\right)
\eq
in eq.~(\ref{def_forward_limit}). 
As we defined the regularised $\loopnumber$-forward limit through non-singular diagrams, multiplying individual diagrams
with a combinatorial factor will not change the considerations in this section and
the regularised $\loopnumber$-forward limit of a tree-amplitude-like object is defined analogously.
}

\section{\sw Symmetry factors, minus signs and combinatorial factors}
\label{sect:symmetry_factors}

In this section we discuss {\sw symmetry factors, minus signs due to closed fermion loops and combinatorial factors}.
We start with the symmetry factors. Let us define the set 
\bq
 {\mathcal U}_{\loopnumber,n}^{\loopnumber-\mathrm{marked}}.
\eq 
Graphs in this set are all graphs which can be obtained from graphs in 
${\mathcal U}_{\loopnumber,n}$ by marking 
a set of $\loopnumber$ internal edges $e_{\sigma_1},...,e_{\sigma_\loopnumber}$ with $1$, ..., $\loopnumber$ and an orientation,
such that when cutting these marked edges we obtain a connected tree graph with $n+2\loopnumber$ external edges.
Graphs in ${\mathcal U}_{\loopnumber,n}^{\loopnumber-\mathrm{marked}}$ are considered to be different, 
if they are obtained from marking different edges
of a graph $\Gamma \in {\mathcal U}_{\loopnumber,n}$.
Furthermore, graphs in ${\mathcal U}_{\loopnumber,n}^{\loopnumber-\mathrm{marked}}$ are considered to be different, if the order of the markings is different or if they differ in the orientation of a marked edge.
In short, elements of ${\mathcal U}_{\loopnumber,n}^{\loopnumber-\mathrm{marked}}$ are graphs with the additional information specified by
an ordered $\loopnumber$-tuple $(e_{\sigma_1},...,e_{\sigma_\loopnumber})$ 
such that cutting the edges $(e_{\sigma_1},...,e_{\sigma_\loopnumber})$ yields
a connected tree graph,
together with a map $e_{\sigma_j} \rightarrow j$, which marks the selected edges with $1$, ..., $\loopnumber$
and a map $e_{\sigma_j} \rightarrow \{+,-\}$, which defines the orientation.
There is a projection
\bq
 \pi_{\mathrm{forget}}
 & : &
 {\mathcal U}_{\loopnumber,n}^{\loopnumber-\mathrm{marked}} 
 \rightarrow
 {\mathcal U}_{\loopnumber,n},
\eq
defined by forgetting the information related to the markings of the internal edges.
Given a graph $\Gamma \in {\mathcal U}_{\loopnumber,n}$ there are
\bq
 2^\loopnumber \loopnumber! \; \left| {\mathcal T}_\Gamma \right|
\eq
graphs in ${\mathcal U}_{\loopnumber,n}^{\loopnumber-\mathrm{marked}}$, which project to $\Gamma$.
This number is easily obtained as follows: $| {\mathcal T}_\Gamma|$ gives all possibilities
a set of markings can be chosen, while the factor $\loopnumber!$ accounts for all possibilities 
of ordering this set and the factor $2^l$ for all possibilities of choosing an orientation.
Thus
\bq
 \left| {\mathcal U}_{\loopnumber,n}^{\loopnumber-\mathrm{marked}} \right|
 & = &
 2^\loopnumber \loopnumber! \; \sum\limits_{\Gamma \in {\mathcal U}_{\loopnumber,n}} 
 \left| {\mathcal T}_\Gamma \right|.
\eq
The set ${\mathcal U}_{\loopnumber,n}^{\mathrm{amputated},\loopnumber-\mathrm{marked}}$ 
is defined analogously as the subset
of graphs $\Gamma \in {\mathcal U}_{\loopnumber,n}^{\loopnumber-\mathrm{marked}}$ 
without self-energy corrections on external lines.

It is also useful to introduce the set 
\bq
 {\mathcal U}_{\loopnumber,n}^{\loopnumber-\mathrm{marked}, \mathrm{\; no \; history}},
\eq
which is obtained 
from ${\mathcal U}_{\loopnumber,n}^{\loopnumber-\mathrm{marked}}$ by forgetting the $n$-tuple $(e_{\sigma_1},...,e_{\sigma_\loopnumber})$,
but keeping the marking with $1$, ..., $l$ and the orientation.
The difference between ${\mathcal U}_{\loopnumber,n}^{\loopnumber-\mathrm{marked}}$ and ${\mathcal U}_{\loopnumber,n}^{\loopnumber-\mathrm{marked}, \mathrm{\; no \; history}}$
is best illustrated by an example.
\begin{figure}
\begin{center}
\includegraphics[scale=1.0]{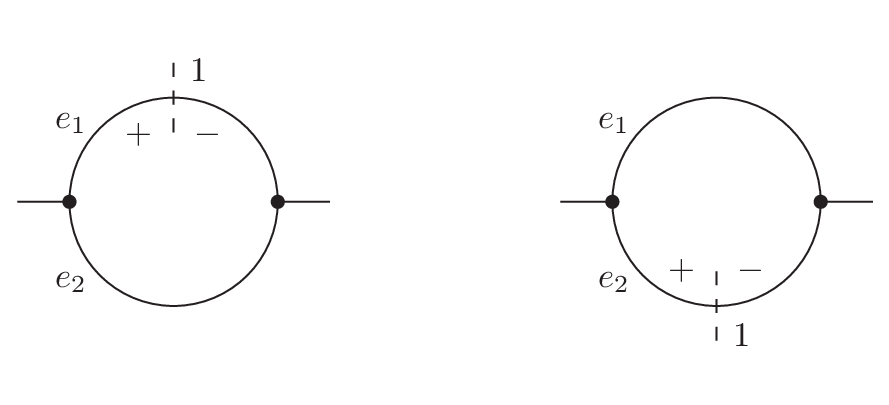}
\end{center}
\caption{
The two graphs shown in this figure correspond to different elements of ${\mathcal U}_{1,2}^{1-\mathrm{marked}}$,
but to the same element of ${\mathcal U}_{1,2}^{1-\mathrm{marked}, \mathrm{\; no \; history}}$.
}
\label{fig_4}
\end{figure}
The two graphs shown in fig.~\ref{fig_4} corresponds to two different elements of ${\mathcal U}_{\loopnumber,n}^{\loopnumber-\mathrm{marked}}$,
but to the same element of ${\mathcal U}_{\loopnumber,n}^{\loopnumber-\mathrm{marked}, \mathrm{\; no \; history}}$.
There are projections
\bq
 \pi_{\mathrm{forget \; history}}
 & : &
 {\mathcal U}_{\loopnumber,n}^{\loopnumber-\mathrm{marked}} 
 \rightarrow
 {\mathcal U}_{\loopnumber,n}^{\loopnumber-\mathrm{marked}, \mathrm{\; no \; history}},
 \nonumber \\
 \pi_{\mathrm{forget \; marking}}
 & : &
 {\mathcal U}_{\loopnumber,n}^{\loopnumber-\mathrm{marked}, \mathrm{\; no \; history}} 
 \rightarrow
 {\mathcal U}_{\loopnumber,n},
\eq
defined in the obvious way, such that $\pi_{\mathrm{forget}} = \pi_{\mathrm{forget \; marking}} \circ \pi_{\mathrm{forget \; history}}$.

Let us consider the set of momenta
\bq
\label{external_momenta_sewed_graph}
 \left\{ p_1, ..., p_n, k_1, ..., k_\loopnumber, \bar{k}_1, ..., \bar{k}_\loopnumber \right\}
\eq
and the set of graphs
\bq
 {\mathcal U}_{0,n+2\loopnumber}^{\loopnumber-\mathrm{sewed}}.
\eq 
Graphs in this set are obtained from tree graphs with $n+2\loopnumber$ external lines labelled by the external momenta
in eq.~(\ref{external_momenta_sewed_graph}), where we sew together the external edges
$k_i$ and $\bar{k}_i$ for all $i \in \{1,...,\loopnumber\}$. 
The sewed edges become internal edges and we obtain a graph with $n$ external edges and $\loopnumber$ loops.
We keep the marking $k_j$ and $\bar{k}_j$ for the sewed half-edges.
The marking of the half-edges with $k_j$ and $\bar{k}_j$ 
defines an orientation of the sewed edges (from $k_j$ to $\bar{k}_j$).
There is a bijection
\bq
 \iota & : &
 {\mathcal U}_{0,n+2\loopnumber}^{\loopnumber-\mathrm{sewed}} 
 \rightarrow
 {\mathcal U}_{\loopnumber,n}^{\loopnumber-\mathrm{marked}, \mathrm{\; no \; history}},
\eq
sending the $j$-th sewed line with label $k_j$ to the orientation label $+$ and the $j$-th sewed line with label $\bar{k}_j$ to the orientation label $-$.

Let us now investigate the relation between ${\mathcal U}_{\loopnumber,n}^{\loopnumber-\mathrm{marked}}$
and ${\mathcal U}_{0,n+2\loopnumber}^{\loopnumber-\mathrm{sewed}}$.
Consider a function $f(\Gamma)$ for $\Gamma \in {\mathcal U}_{\loopnumber,n}^{\loopnumber-\mathrm{marked}, \mathrm{\; no \; history}}$.
We have
\bq
\label{graph_summation}
 \sum\limits_{\Gamma \in {\mathcal U}_{\loopnumber,n}^{\loopnumber-\mathrm{marked}}}
 \frac{1}{\left|\mathrm{Aut}\left(\pi_{\mathrm{forget}}\left(\Gamma\right)\right)\right|} 
 f\left(\pi_{\mathrm{forget \; history}}\left(\Gamma\right)\right)
 & = &
 \sum\limits_{\Gamma \in {\mathcal U}_{0,n+2\loopnumber}^{\loopnumber-\mathrm{sewed}}}
 f\left(\iota\left(\Gamma\right)\right).
 \nonumber \\
\eq
Eq.~(\ref{graph_summation}) allows us to replace a summation over ${\mathcal U}_{\loopnumber,n}^{\loopnumber-\mathrm{marked}}$
with symmetry factors inherited from ${\mathcal U}_{\loopnumber,n}$ by
a summation over ${\mathcal U}_{0,n+2\loopnumber}^{\loopnumber-\mathrm{sewed}}$.
{\sw
Eq.~(\ref{graph_summation}) holds for any function $f(\Gamma)$.
In particular, eq.~(\ref{graph_summation}) holds for the case where $f(\Gamma)$ is given as the product of the standard application
of the Feynman rules and the combinatorial factor $S_{\sigma \alpha}$.
}

Before giving a proof of eq.~(\ref{graph_summation}), let us first consider a few examples:
Consider first a graph $\Gamma \in {\mathcal U}_{\loopnumber,n}^{\loopnumber-\mathrm{marked}}$, such that the underlying graph
in ${\mathcal U}_{\loopnumber,n}$ has a trivial symmetry factor, i.e.
\bq
 \left|\mathrm{Aut}\left(\pi_{\mathrm{forget}}\left(\Gamma\right)\right)\right|
 & = &
 1.
\eq
The set $\iota^{-1}\left(\pi_{\mathrm{forget \; history}}\left(\Gamma\right)\right) \subset {\mathcal U}_{0,n+2\loopnumber}^{\loopnumber-\mathrm{sewed}}$ consists of a single graphs, and both sides of eq.~(\ref{graph_summation}) match trivially.
\begin{figure}
\begin{center}
\includegraphics[scale=1.0]{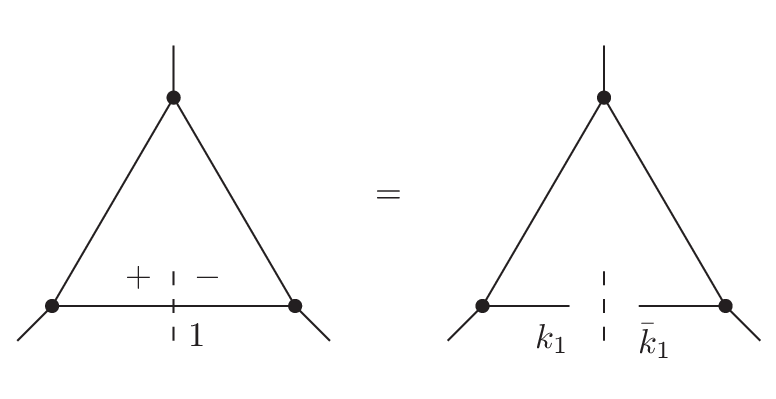}
\end{center}
\caption{
The left-hand side shows a graph $\Gamma \in {\mathcal U}_{1,3}^{1-\mathrm{marked}}$, 
the right-hand side the corresponding set $\iota^{-1}\left(\pi_{\mathrm{forget \; history}}\left(\Gamma\right)\right) \subset {\mathcal U}_{0,5}^{1-\mathrm{sewed}}$.
}
\label{fig_1}
\end{figure}
This is illustrated for a one-loop three-point function in fig.~\ref{fig_1}.

\begin{figure}
\begin{center}
\includegraphics[scale=1.0]{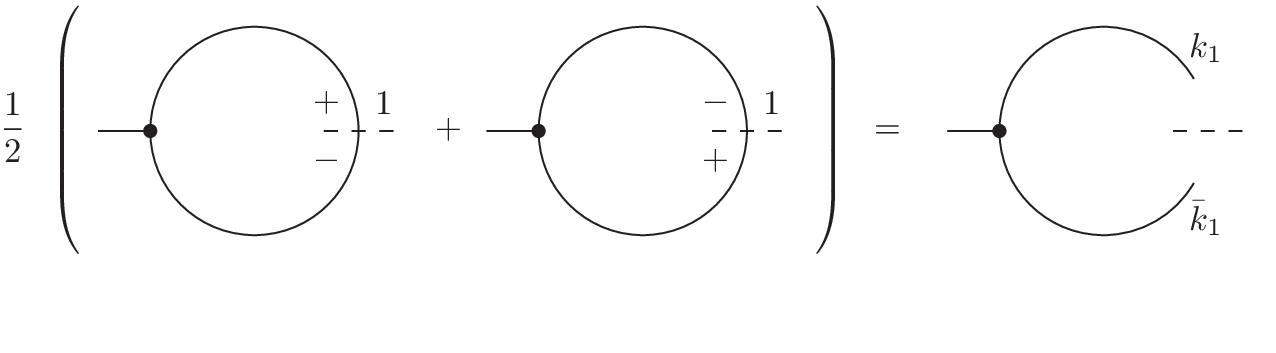}
\end{center}
\caption{
The left-hand side shows two graphs $\Gamma_1, \Gamma_2 \in {\mathcal U}_{1,1}^{1-\mathrm{marked}}$ together with the symmetry factor $1/2$.
We have $\pi_{\mathrm{forget \; history}}\left(\Gamma_1\right) = \pi_{\mathrm{forget \; history}}\left(\Gamma_2\right) = \tilde{\Gamma}$.
The right-hand side shows the corresponding set $\iota^{-1}\left(\tilde{\Gamma}\right) \subset {\mathcal U}_{0,3}^{1-\mathrm{sewed}}$.
}
\label{fig_2}
\end{figure}

Let us now discuss the case of non-trivial symmetry factors.
We start with two one-loop one-point graph $\Gamma_1, \Gamma_2 \in {\mathcal U}_{1,1}^{1-\mathrm{marked}}$ with one marking, 
as shown in fig.~\ref{fig_2}.
Since
\bq
 \left|\mathrm{Aut}\left(\pi_{\mathrm{forget}}\left(\Gamma_1\right)\right)\right|
 \;\; = \;\;
 \left|\mathrm{Aut}\left(\pi_{\mathrm{forget}}\left(\Gamma_2\right)\right)\right|
 \;\; = \;\;
 2,
\eq
both graphs inherit a symmetry factor $1/2$.
The two graphs $\Gamma_1$ and $\Gamma_2$ differ only in the orientation of the marked edge, but project to the same graph
$\tilde{\Gamma} \in {\mathcal U}_{1,1}^{1-\mathrm{marked}, \mathrm{\; no \; history}}$:
\bq
  \tilde{\Gamma} \; = \; \pi_{\mathrm{forget \; history}}\left(\Gamma_1\right) \; = \; \pi_{\mathrm{forget \; history}}\left(\Gamma_2\right)
\eq
If we exchange the orientation $+$ with $-$ we obtain the same unordered graph in
${\mathcal U}_{1,1}^{1-\mathrm{marked}, \mathrm{\; no \; history}}$.
Therefore we have
\bq
 f\left(\pi_{\mathrm{forget \; history}}\left(\Gamma_1\right)\right)
 & = &
 f\left(\pi_{\mathrm{forget \; history}}\left(\Gamma_2\right)\right).
\eq
The set $\iota^{-1}\left(\tilde{\Gamma}\right) \subset {\mathcal U}_{0,3}^{1-\mathrm{sewed}}$ consists of
one graph.
Thus the symmetry factor $1/2$ on the left-hand side of eq.~(\ref{graph_summation}) cancels the over-counting in 
${\mathcal U}_{1,1}^{1-\mathrm{marked}}$.

\begin{figure}
\begin{center}
\includegraphics[scale=0.88]{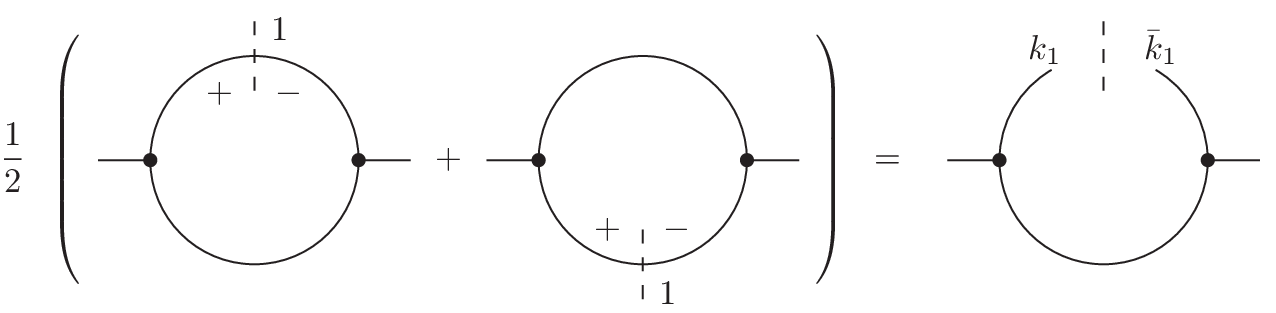}
\end{center}
\caption{
The left-hand side shows two graphs $\Gamma_1, \Gamma_2 \in {\mathcal U}_{1,2}^{1-\mathrm{marked}}$ together with the symmetry factor $1/2$.
We have $\pi_{\mathrm{forget \; history}}\left(\Gamma_1\right) = \pi_{\mathrm{forget \; history}}\left(\Gamma_2\right) = \tilde{\Gamma}$.
The right-hand side shows the corresponding set $\iota^{-1}\left(\tilde{\Gamma}\right) \subset {\mathcal U}_{0,4}^{1-\mathrm{sewed}}$.
}
\label{fig_3}
\end{figure}

As a third example consider two one-loop two-point graphs $\Gamma_1, \Gamma_2 \in  {\mathcal U}_{1,2}^{1-\mathrm{marked}}$
with one marking, as shown in fig.\ref{fig_3}.
The two graphs $\Gamma_1$ and $\Gamma_2$ differ only in the choice of the marked edge, but project to the same graph
$\tilde{\Gamma} \in {\mathcal U}_{1,2}^{1-\mathrm{marked}, \mathrm{\; no \; history}}$.
On the side of the sewed graphs, the set $\iota^{-1}\left(\tilde{\Gamma}\right) \subset {\mathcal U}_{0,4}^{1-\mathrm{sewed}}$ consists of
one graph.
Thus the symmetry factor $1/2$ on the left-hand side of eq.~(\ref{graph_summation}) cancels the over-counting in 
${\mathcal U}_{1,2}^{1-\mathrm{marked}}$.

\begin{figure}
\begin{center}
\includegraphics[scale=0.88]{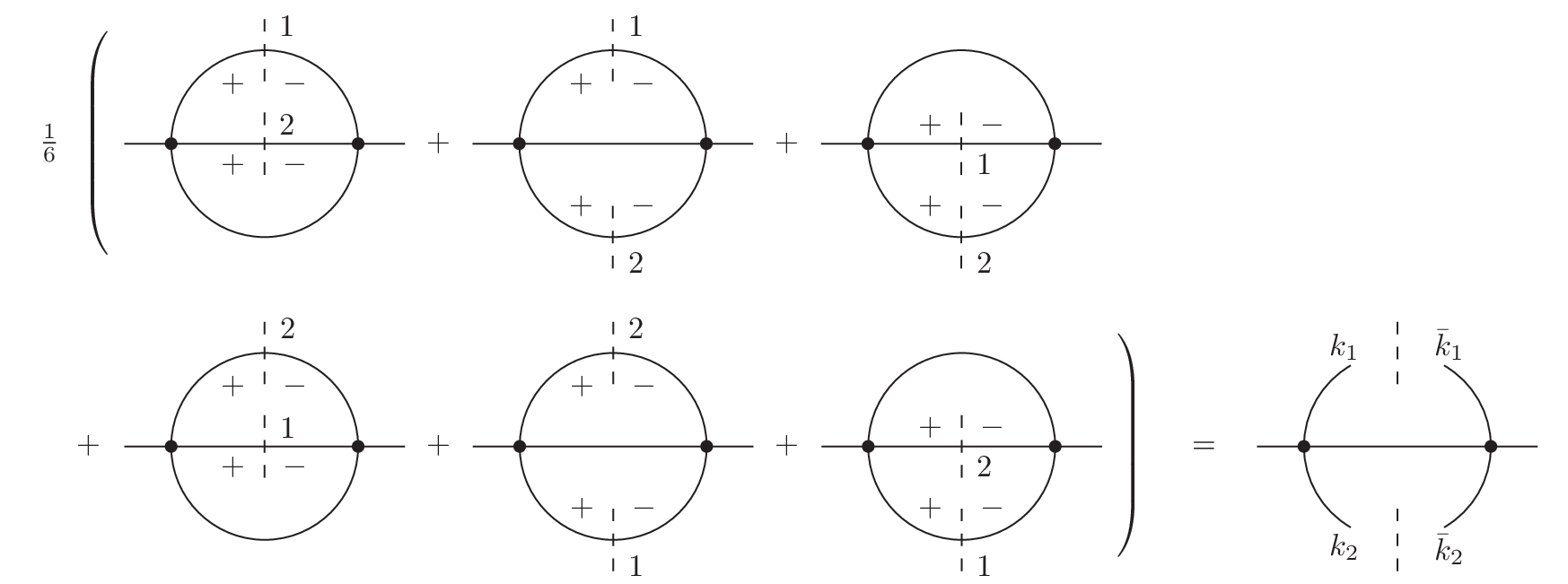}
\end{center}
\caption{
The left-hand side shows six graphs $\Gamma_1, ..., \Gamma_6 \in {\mathcal U}_{2,2}^{2-\mathrm{marked}}$ together with the symmetry factor $1/6$.
We have $\pi_{\mathrm{forget \; history}}\left(\Gamma_i\right) = \tilde{\Gamma}$ for $i=1,...,6$.
The right-hand side shows the corresponding set $\iota^{-1}\left(\tilde{\Gamma}\right) \subset {\mathcal U}_{0,6}^{2-\mathrm{sewed}}$.
}
\label{fig_20}
\end{figure}

As a final example we consider six two-loop two-point graphs $\Gamma_1, ..., \Gamma_6 \in  {\mathcal U}_{2,2}^{2-\mathrm{marked}}$
with two markings, as shown in fig.\ref{fig_20}.
The six graphs $\Gamma_1, ..., \Gamma_6$ differ only in the choice of the marked edges, but project to the same graph
$\tilde{\Gamma} \in {\mathcal U}_{2,2}^{2-\mathrm{marked}, \mathrm{\; no \; history}}$.
On the side of the sewed graphs, the set $\iota^{-1}\left(\tilde{\Gamma}\right) \subset {\mathcal U}_{0,6}^{2-\mathrm{sewed}}$ consists of
one graph.
Thus the symmetry factor $1/6$ on the left-hand side of eq.~(\ref{graph_summation}) cancels the over-counting in 
${\mathcal U}_{2,2}^{2-\mathrm{marked}}$.

Let us now give a proof of eq.~(\ref{graph_summation}).
We have already seen that there is a bijection between the graphs in
${\mathcal U}_{0,n+2\loopnumber}^{\loopnumber-\mathrm{sewed}}$ and ${\mathcal U}_{\loopnumber,n}^{\loopnumber-\mathrm{marked}, \mathrm{\; no \; history}}$.
On the other hand, there might be several graphs in ${\mathcal U}_{\loopnumber,n}^{\loopnumber-\mathrm{marked}}$
which project to the same graph in ${\mathcal U}_{\loopnumber,n}^{\loopnumber-\mathrm{marked}, \mathrm{\; no \; history}}$.
We have to show that the symmetry factor exactly compensates this over-counting.
Let us consider a graph $\Gamma \in {\mathcal U}_{\loopnumber,n}^{\loopnumber-\mathrm{marked}}$.
An automorphism $T \in \mathrm{Aut}\left(\pi_{\mathrm{forget}}\left(\Gamma\right)\right)$ permutes the edges and vertices
of $\pi_{\mathrm{forget}}\left(\Gamma\right)$ and induces a group action on ${\mathcal U}_{\loopnumber,n}^{\loopnumber-\mathrm{marked}}$
by permuting the corresponding edges and vertices together with the markings and the orientation.
Let us look at the orbit of $\Gamma$ under the group action.
It is clear that all graphs in the orbit of $\Gamma$ project to the same graph $\pi_{\mathrm{forget \; history}}\left(\Gamma\right)$
in ${\mathcal U}_{\loopnumber,n}^{\loopnumber-\mathrm{marked}, \mathrm{\; no \; history}}$.
On the other hand a graph $\Gamma \in {\mathcal U}_{\loopnumber,n}^{\loopnumber-\mathrm{marked}}$ corresponds (due to the markings) 
to a tree graph,
therefore its automorphism group is trivial.
Hence the stabiliser group of the group action is trivial and the group action is free.
The orbit of the induced action of $\mathrm{Aut}\left(\pi_{\mathrm{forget}}\left(\Gamma\right)\right)$
of $\Gamma$ generates all graphs which project to $\pi_{\mathrm{forget \; history}}\left(\Gamma\right)$
and the symmetry factor correctly compensates the over-counting.

Let us now discuss the minus signs for each closed fermion loop.
We recall that we defined in section~\ref{sect:cutting_sewing}
the sewing operation in such a way that it includes a minus sign for each sewing of a fermion loop.
It is immediately clear that this prescription reproduces the required additional minus sign for each closed
fermion loop.
However, what is not immediately obvious is how this minus sign cancels with another minus sign in the case where
the sewing operation does not lead to a closed fermion loop.
In order to see the mechanism, we have to discuss tree amplitudes with fermion-antifermion pairs of identical flavour.
These amplitudes can always be related to amplitudes,
where all fermion-antifermion pairs have different flavours.
This is achieved by summing over all fermion permutations.
An amplitude with $n_f$ identical fermion-antifermion pairs can be written as
\bq
\label{identical_fermions}
\lefteqn{
{\mathcal A}_{0,n}\left( \bar{f}_1, f_1, ..., \bar{f}_2, f_2, ..., \bar{f}_{n_f}, f_{n_f} \right)
 = } & & \nonumber \\
 & & 
 \sum\limits_{\sigma \in S(n_f)} \left( -1 \right)^{\sigma} 
     {\mathcal A}_{0,n}^{\mathrm{non-id}}\left( \bar{f}_1, f_{\sigma(1)}, ..., \bar{f}_2, f_{\sigma(2)}, ..., \bar{f}_{n_f}, f_{\sigma(n_f)} \right).
\eq
Here, $(-1)^\sigma$ equals $-1$ whenever the permutation is odd and equals $+1$ if the permutation is even.
In ${\mathcal A}_{0,n}^{\mathrm{non-id}}$ each external fermion-antifermion pair $(\bar{f}_j, f_{\sigma(j)})$ is connected 
by a continuous fermion line and treated as having a flavour different from all other
fermion-antifermion pairs.
To give an example
\bq
\label{fermion_example}
 {\mathcal A}_{0,4}\left( \bar{f}_1, f_1, \bar{f}_2, f_2 \right)
 & = &
 {\mathcal A}_{0,4}^{\mathrm{non-id}}\left( \bar{f}_1, f_1, \bar{f}_2, f_2 \right)
 -
 {\mathcal A}_{0,4}^{\mathrm{non-id}}\left( \bar{f}_1, f_2, \bar{f}_2, f_1 \right).
\eq
There is only one diagram contributing to ${\mathcal A}_{0,4}^{\mathrm{non-id}}( \bar{f}_1, f_1, \bar{f}_2, f_2)$
and ${\mathcal A}_{0,4}^{\mathrm{non-id}}( \bar{f}_1, f_2, \bar{f}_2, f_1)$.
Thus ${\mathcal A}_{0,4}( \bar{f}_1, f_1, \bar{f}_2, f_2)$ is the sum of two Feynman diagrams.
\begin{figure}
\begin{center}
\includegraphics[scale=1.0]{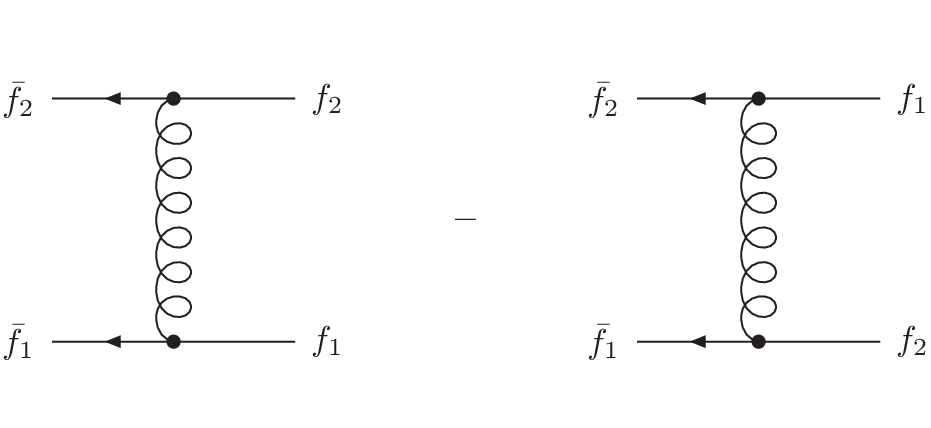}
\end{center}
\caption{
The two diagrams contributing to ${\mathcal A}_{0,4}( \bar{f}_1, f_1, \bar{f}_2, f_2 )$.
}
\label{fig_17}
\end{figure}
This is shown in fig.~\ref{fig_17}.
Let us now sew $\bar{f}_2$ with $f_2$.
Including the minus sign from the sewing operation, the first term 
on the right-hand side of eq.~(\ref{fermion_example}) gives us minus the tadpole with a closed fermion loop,
while the second term gives us the fermion self-energy with the correct plus sign.
\begin{figure}
\begin{center}
\includegraphics[scale=1.0]{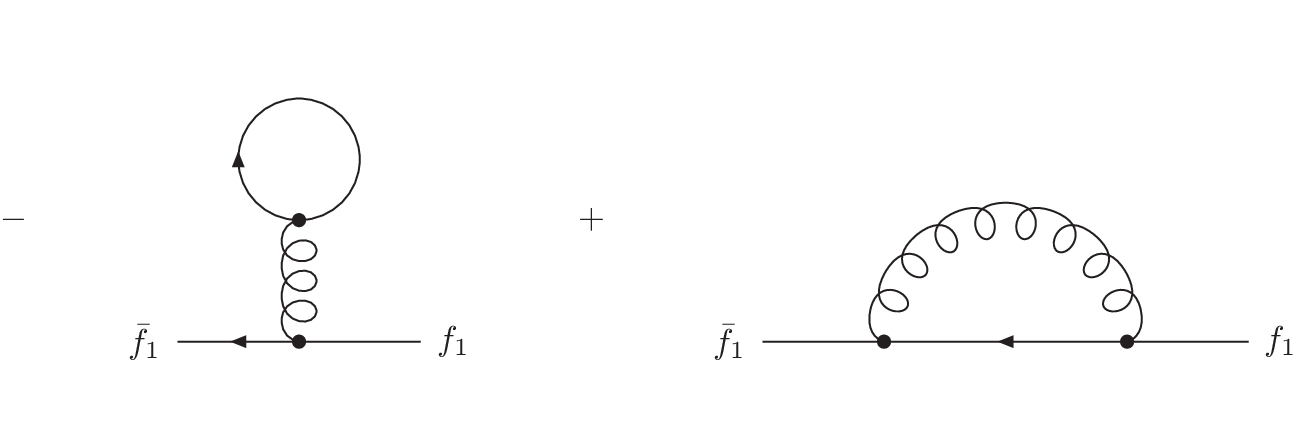}
\end{center}
\caption{
Sewing $f_2$ with $\bar{f}_2$ in the two diagrams of ${\mathcal A}_{0,4}( \bar{f}_1, f_1, \bar{f}_2, f_2 )$
gives the tadpole diagram with a minus sign and the self-energy diagram with a plus sign.
}
\label{fig_18}
\end{figure}
This is shown in fig.~\ref{fig_18}.
We see that the minus sign from the sewing operation cancels with a minus for an odd permutation in eq.~(\ref{identical_fermions}).

The same considerations apply to Faddeev-Popov ghosts:
Ghost lines are treated as fermion lines.

{\sw
Let us now discuss the combinatorial factor $S_{\sigma \alpha}$.
By contruction, the combinatorial factor $S_{\sigma \alpha}$ depends only on the underlying chain graph $\Gamma^{\mathrm{chain}}$.
\begin{figure}
\begin{center}
\includegraphics[scale=1.0]{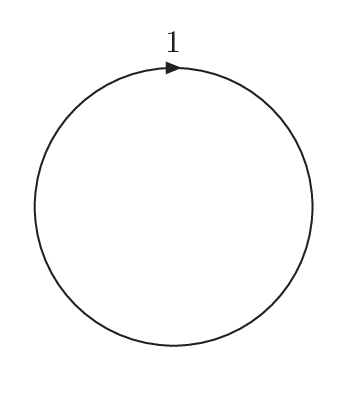}
\hspace*{20mm}
\includegraphics[scale=1.0]{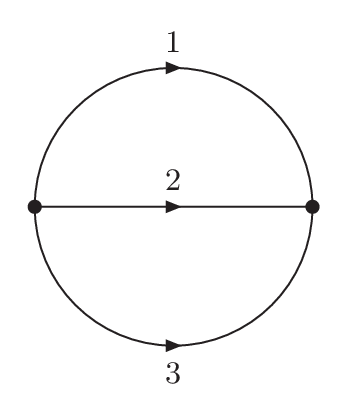}
\hspace*{20mm}
\includegraphics[scale=1.0]{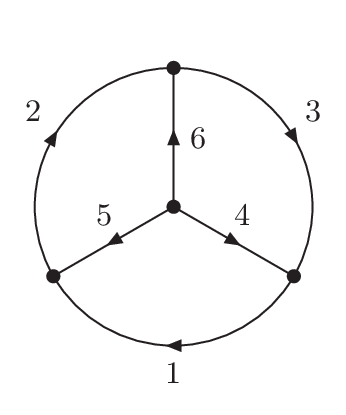}
\end{center}
\caption{
The basic chain graphs up to three loops. Up to this loop order, all other chain graphs are subtopologies of these
three graphs.
}
\label{fig_21}
\end{figure}
Up to three loops, all chain graphs are (sub-) topologies of the three chain graphs shown in fig.~(\ref{fig_21}).
\begin{figure}
\begin{center}
\includegraphics[scale=1.0]{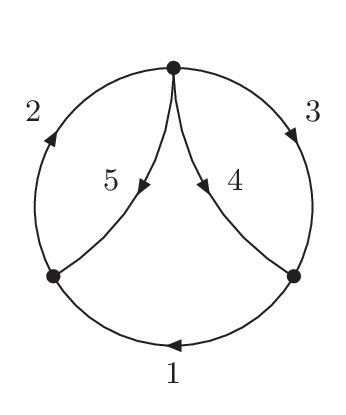}
\hspace*{20mm}
\includegraphics[scale=1.0]{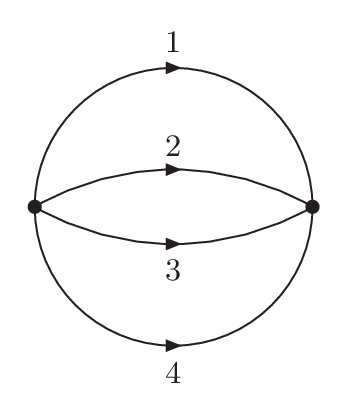}
\end{center}
\caption{
The two non-factorisable subtopologies of the three-loop Mercedes-Benz graph.
}
\label{fig_24}
\end{figure}
The two non-factorisable subtopologies of the three-loop Mercedes-Benz graph are shown in fig.~(\ref{fig_24}).
Up to three loop there aren't  too many chain graphs and we may compute the combinatororial factor for these
graphs once and for all.
To specify a cut, we write
\bq
 \left( \sigma_1^{\alpha_1}, ..., \sigma_\loopnumber^{\alpha_\loopnumber} \right)
\eq
The combinatorial factors are invariant if we change the energy signs $\alpha_j \rightarrow - \alpha_j$ 
for all $j \in \{1,...,\loopnumber\}$.
One finds for the three chain diagrams shown in fig.~(\ref{fig_21})
\bq
\label{combinatorial_factors_I}
\mbox{One-loop:}
 & &
\begin{tabular}{c|c}
 $\mathrm{Cut}$ & $(1^+)$ \\
\hline
 $S_{\sigma \alpha}$ & $\frac{1}{2}$ \\
\end{tabular}
\nonumber \\
\mbox{Two-loop:}
 & &
\begin{tabular}{c|c|c}
 $\mathrm{Cut}$ & $(1^+,2^+)$ & $(1^+,2^-)$ \\
\hline
 $S_{\sigma \alpha}$ & $\frac{1}{3}$ & $\frac{1}{6}$ \\
\end{tabular}
\nonumber \\
\mbox{Three-loop:}
 & &
\begin{tabular}{c|c|c|c|c}
 $\mathrm{Cut}$ & $(1^+,2^+,3^+)$ & $(1^+,2^+,3^-)$ & $(1^+,2^-,3^+)$ & $(1^+,2^-,3^-)$ \\
\hline
 $S_{\sigma \alpha}$ & $\frac{3}{64}$ & $\frac{29}{192}$ & $\frac{29}{192}$ & $\frac{29}{192}$ \\
\end{tabular}
 \nonumber \\
 & &
\begin{tabular}{c|c|c|c|c}
 $\mathrm{Cut}$ & $(1^+,2^+,4^+)$ & $(1^+,2^+,4^-)$ & $(1^+,2^-,4^+)$ & $(1^+,2^-,4^-)$ \\
\hline
 $S_{\sigma \alpha}$ & $\frac{5}{96}$ & $\frac{19}{192}$ & $\frac{19}{192}$ & $\frac{1}{4}$ \\
\end{tabular}
\eq
For the two non-factorisable subtopologies of the Mercedes-Benz graph, shown in fig.~(\ref{fig_24}), one finds
\bq
\label{combinatorial_factors_II}
\mbox{Five propagator graph:}
 & &
\begin{tabular}{c|c|c|c|c}
 $\mathrm{Cut}$ & $(3^+,4^+,5^+)$ & $(3^+,4^+,5^-)$ & $(3^+,4^-,5^+)$ & $(3^+,4^-,5^-)$ \\
\hline
 $S_{\sigma \alpha}$ & $\frac{1}{4}$ & $\frac{11}{96}$ & $\frac{13}{192}$ & $\frac{13}{192}$ \\
\end{tabular}
 \nonumber \\
 & &
\begin{tabular}{c|c|c|c|c}
 $\mathrm{Cut}$ & $(1^+,3^+,5^+)$ & $(1^+,3^+,5^-)$ & $(1^+,3^-,5^+)$ & $(1^+,3^-,5^-)$ \\
\hline
 $S_{\sigma \alpha}$ & $\frac{1}{8}$ & $\frac{11}{192}$ & $\frac{37}{192}$ & $\frac{1}{8}$ \\
\end{tabular}
 \nonumber \\
\mbox{Four propagator graph:}
 & &
\begin{tabular}{c|c|c|c|c}
 $\mathrm{Cut}$ & $(1^+,2^+,3^+)$ & $(1^+,2^+,3^-)$ & $(1^+,2^-,3^+)$ & $(1^+,2^-,3^-)$ \\
\hline
 $S_{\sigma \alpha}$ & $\frac{1}{4}$ & $\frac{1}{12}$ & $\frac{1}{12}$ & $\frac{1}{12}$ \\
\end{tabular}
 \nonumber \\
\eq
Up to three loops one easily verifies that one 
may distribute the combinatorial factor over the vertices and propagators of the chain diagram, such
that the combinatorial factor is the product of all vertex and propagator factors.
From the three chain diagrams shown in fig.~(\ref{fig_21}) we deduce the vertex factors
\begin{alignat}{3}
\begin{picture}(90,60)(0,40)
 \Line(30,50)(30,70)
 \Line(70,50)(70,70)
 \Line(50,30)(30,50)
 \Line(50,30)(70,50)
 \Line(50,10)(50,30)
 \GCirc(30,50){10}{0.5}
 \GCirc(70,50){10}{0.5}
 \Text(30,75)[b]{\small $k_1^+$}
 \Text(70,75)[b]{\small $\bar{k}_1^-$}
 \SetColor{Red}
 \Vertex(50,30){2}
 \SetColor{Black}
\end{picture}
 & \;\; = \;\; \frac{1}{2},
 & & & & \nonumber \\
 & & & & & \nonumber \\
\begin{picture}(90,60)(0,40)
 \Line(30,50)(30,70)
 \Line(70,50)(70,70)
 \Line(50,30)(30,50)
 \Line(50,30)(70,50)
 \Line(50,10)(50,30)
 \GCirc(30,50){10}{0.5}
 \GCirc(70,50){10}{0.5}
 \Text(30,75)[b]{\small $k_1^+$}
 \Text(70,75)[b]{\small $k_2^+$}
 \SetColor{Red}
 \Vertex(50,30){2}
 \SetColor{Black}
\end{picture}
 & \;\; = \;\; \frac{1}{\sqrt{3}}, 
 & 
\begin{picture}(90,60)(0,40)
 \Line(30,50)(30,70)
 \Line(70,50)(70,70)
 \Line(50,30)(30,50)
 \Line(50,30)(70,50)
 \Line(50,10)(50,30)
 \GCirc(30,50){10}{0.5}
 \GCirc(70,50){10}{0.5}
 \Text(30,75)[b]{\small $k_1^+$}
 \Text(70,75)[b]{\small $k_2^-$}
 \SetColor{Red}
 \Vertex(50,30){2}
 \SetColor{Black}
\end{picture}
 & \;\; = \;\; \frac{1}{\sqrt{6}}, & & \nonumber \\
 & & & & & \nonumber \\
 & & & & & \nonumber \\
\begin{picture}(90,60)(0,40)
 \Line(22.5,50)(22.5,70)
 \Line(37.5,50)(37.5,70)
 \Line(70,50)(70,70)
 \Line(50,30)(30,50)
 \Line(50,30)(70,50)
 \Line(50,10)(50,30)
 \GOval(30,50)(10,12.5)(0){0.5}
 \GCirc(70,50){10}{0.5}
 \Text(22.5,75)[b]{\small $k_1^+$}
 \Text(37.5,75)[b]{\small $k_2^+$}
 \Text(70,75)[b]{\small $k_3^+$}
 \SetColor{Red}
 \Vertex(50,30){2}
 \SetColor{Black}
\end{picture}
 & \;\; = \;\; \frac{1}{2} \sqrt{3}, 
 & 
\begin{picture}(90,60)(0,40)
 \Line(22.5,50)(22.5,70)
 \Line(37.5,50)(37.5,70)
 \Line(70,50)(70,70)
 \Line(50,30)(30,50)
 \Line(50,30)(70,50)
 \Line(50,10)(50,30)
 \GOval(30,50)(10,12.5)(0){0.5}
 \GCirc(70,50){10}{0.5}
 \Text(22.5,75)[b]{\small $k_1^+$}
 \Text(37.5,75)[b]{\small $k_2^-$}
 \Text(70,75)[b]{\small $k_3^+$}
 \SetColor{Red}
 \Vertex(50,30){2}
 \SetColor{Black}
\end{picture}
 & \;\; = \;\; \frac{1}{4} \sqrt{5},
 & 
\begin{picture}(90,60)(0,40)
 \Line(22.5,50)(22.5,70)
 \Line(37.5,50)(37.5,70)
 \Line(70,50)(70,70)
 \Line(50,30)(30,50)
 \Line(50,30)(70,50)
 \Line(50,10)(50,30)
 \GOval(30,50)(10,12.5)(0){0.5}
 \GCirc(70,50){10}{0.5}
 \Text(22.5,75)[b]{\small $k_1^+$}
 \Text(37.5,75)[b]{\small $k_2^+$}
 \Text(70,75)[b]{\small $k_3^-$}
 \SetColor{Red}
 \Vertex(50,30){2}
 \SetColor{Black}
\end{picture}
 & \;\; = \;\; \frac{19}{80} \sqrt{10}, \nonumber \\
 & & & & & \nonumber \\
\begin{picture}(90,60)(0,40)
 \Line(22.5,50)(22.5,70)
 \Line(37.5,50)(37.5,70)
 \Line(62.5,50)(62.5,70)
 \Line(77.5,50)(77.5,70)
 \Line(50,30)(30,50)
 \Line(50,30)(70,50)
 \Line(50,10)(50,30)
 \GOval(30,50)(10,12.5)(0){0.5}
 \GOval(70,50)(10,12.5)(0){0.5}
 \Text(22.5,75)[b]{\small $k_1^+$}
 \Text(37.5,75)[b]{\small $k_2^+$}
 \Text(62.5,75)[b]{\small $\bar{k}_2^-$}
 \Text(77.5,75)[b]{\small $k_3^-$}
 \SetColor{Red}
 \Vertex(50,30){2}
 \SetColor{Black}
\end{picture}
 & \;\; = \;\; \frac{29}{64} \sqrt{6}, 
 & 
\begin{picture}(90,60)(0,40)
 \Line(22.5,50)(22.5,70)
 \Line(37.5,50)(37.5,70)
 \Line(62.5,50)(62.5,70)
 \Line(77.5,50)(77.5,70)
 \Line(50,30)(30,50)
 \Line(50,30)(70,50)
 \Line(50,10)(50,30)
 \GOval(30,50)(10,12.5)(0){0.5}
 \GOval(70,50)(10,12.5)(0){0.5}
 \Text(22.5,75)[b]{\small $k_1^+$}
 \Text(37.5,75)[b]{\small $k_2^-$}
 \Text(62.5,75)[b]{\small $\bar{k}_2^+$}
 \Text(77.5,75)[b]{\small $k_3^-$}
 \SetColor{Red}
 \Vertex(50,30){2}
 \SetColor{Black}
\end{picture}
 & \;\; = \;\; \frac{9}{32} \sqrt{6}.
 & & \nonumber \\
\end{alignat}
These vertex factors are sufficient to reproduce the correct combinatorial factor for any
graph obtained from dressing up the the chain graphs in fig.~(\ref{fig_21}) with external lines.
The propagator factors are all $1$ in this case.
In order to see how it works, it is best to give an example. We consider the cut
$(1^+,2^+,3^+)$ of the Mercedes-Benz graph.  
\begin{figure}
\begin{center}
\includegraphics[scale=1.0]{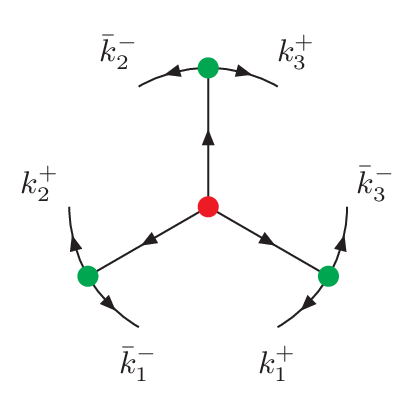}
\end{center}
\caption{
The cut $(1^+,2^+,3^+)$ of the Mercedes-Benz graph.
}
\label{fig_27}
\end{figure}
This cut is shown in fig.~(\ref{fig_27}).
In fig.~(\ref{fig_27}) three vertices are shown in green, one vertex in red.
Each green vertex contributes a factor $1/\sqrt{6}$, while the red vertex contributes a factor
$9 \sqrt{6}/32$.
In total we obtain
\bq
 \left( \frac{1}{\sqrt{6}} \right)
 \left( \frac{1}{\sqrt{6}} \right)
 \left( \frac{1}{\sqrt{6}} \right)
 \left( \frac{9}{32} \sqrt{6} \right)
 & = &
 \frac{3}{64},
\eq
which is the combinatorial factor for this cut.

The two subtopologies of the Mercedes-Benz graph shown in fig.~(\ref{fig_24}) involve vertices with valency $4$.
They give new vertex factors for vertices of valency $4$.
We are primarly interested in the application of the methods to QCD.
It is well-known, that with the help of an auxiliary tensor particle we may eliminate the four-gluon vertex, 
such that we have to deal with three-valent vertices only.
It may seem that the vertex factors for vertices of valency $3$ are all what is needed.
However, this is not quite true. The reason is the following:
\begin{figure}
\begin{center}
\includegraphics[scale=1.0]{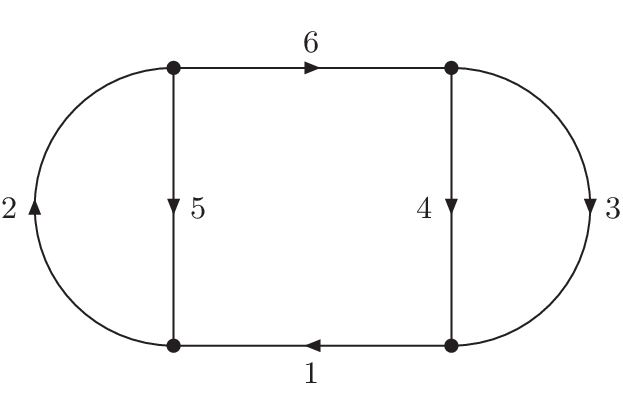}
\end{center}
\caption{
A three-loop graph. Propagators $1$ and $2$ belong to the same chain. 
The underlying chain graph is the five-propagator graph shown in fig.~(\ref{fig_24}).
}
\label{fig_26}
\end{figure}
At three-loops there is the ladder graph shown in fig.~(\ref{fig_26}), involving only vertices of valency $3$.
This graph is not a chain graph, as the same loop momentum is flowing through the propagators $1$ and $6$.
The underlying chain graph is the five-propagator graph shown in fig.~(\ref{fig_24}).
The three-loop ladder graph has two types of cuts, representeted by $(3,4,5)$ and $(1,3,5)$.
The vertex factors defined up to now don't reproduce the correct combinatorial factor 
(given by the combinatorial factor of the five-propagator graph), and we introduce
for this graph non-trivial propagator factors.
For cuts of the type $(3,4,5)$ the two propagators $1$ and $6$ are un-cut.
For one of the two (but not both) we introduce a non-trivial propagator factor according to
\begin{alignat}{2}
\label{comb_factor_second_prop}
\begin{picture}(140,60)(0,40)
 \Line(30,50)(30,70)
 \Line(70,50)(70,70)
 \Line(110,50)(110,70)
 \Line(30,50)(50,30)
 \Line(50,30)(70,50)
 \Line(110,50)(70,10)
 \Line(70,10)(70,-10)
 \GCirc(30,50){10}{0.5}
 \GCirc(70,50){10}{0.5}
 \GCirc(110,50){10}{0.5}
 \Text(30,75)[b]{\small $k_1^+$}
 \Text(70,75)[b]{\small $k_2^+$}
 \Text(110,75)[b]{\small $k_3^+$}
 \SetColor{Red}
 \SetWidth{2.0}
 \Line(50,30)(70,10)
 \SetWidth{0.5}
 \SetColor{Black}
 \Vertex(50,30){2}
 \Vertex(70,10){2}
\end{picture}
 & \;\; = \;\; 1,
 & 
\begin{picture}(140,60)(0,40)
 \Line(30,50)(30,70)
 \Line(70,50)(70,70)
 \Line(110,50)(110,70)
 \Line(30,50)(50,30)
 \Line(50,30)(70,50)
 \Line(110,50)(70,10)
 \Line(70,10)(70,-10)
 \GCirc(30,50){10}{0.5}
 \GCirc(70,50){10}{0.5}
 \GCirc(110,50){10}{0.5}
 \Text(30,75)[b]{\small $k_1^+$}
 \Text(70,75)[b]{\small $k_2^+$}
 \Text(110,75)[b]{\small $k_3^-$}
 \SetColor{Red}
 \SetWidth{2.0}
 \Line(50,30)(70,10)
 \SetWidth{0.5}
 \SetColor{Black}
 \Vertex(50,30){2}
 \Vertex(70,10){2}
\end{picture}
 & \;\; = \;\; \frac{220}{361}, \nonumber \\
 & & & \nonumber \\
 & & & \nonumber \\
\begin{picture}(140,60)(0,40)
 \Line(30,50)(30,70)
 \Line(70,50)(70,70)
 \Line(110,50)(110,70)
 \Line(30,50)(50,30)
 \Line(50,30)(70,50)
 \Line(110,50)(70,10)
 \Line(70,10)(70,-10)
 \GCirc(30,50){10}{0.5}
 \GCirc(70,50){10}{0.5}
 \GCirc(110,50){10}{0.5}
 \Text(30,75)[b]{\small $k_1^+$}
 \Text(70,75)[b]{\small $k_2^-$}
 \Text(110,75)[b]{\small $k_3^+$}
 \SetColor{Red}
 \SetWidth{2.0}
 \Line(50,30)(70,10)
 \SetWidth{0.5}
 \SetColor{Black}
 \Vertex(50,30){2}
 \Vertex(70,10){2}
\end{picture}
 & \;\; = \;\; \frac{13}{10}, 
 & 
\begin{picture}(140,60)(0,40)
 \Line(30,50)(30,70)
 \Line(70,50)(70,70)
 \Line(110,50)(110,70)
 \Line(30,50)(50,30)
 \Line(50,30)(70,50)
 \Line(110,50)(70,10)
 \Line(70,10)(70,-10)
 \GCirc(30,50){10}{0.5}
 \GCirc(70,50){10}{0.5}
 \GCirc(110,50){10}{0.5}
 \Text(30,75)[b]{\small $k_1^+$}
 \Text(70,75)[b]{\small $k_2^-$}
 \Text(110,75)[b]{\small $k_3^-$}
 \SetColor{Red}
 \SetWidth{2.0}
 \Line(50,30)(70,10)
 \SetWidth{0.5}
 \SetColor{Black}
 \Vertex(50,30){2}
 \Vertex(70,10){2}
\end{picture}
 & \;\; = \;\; \frac{13}{10}. \nonumber \\
 & & & \nonumber \\
 & & & \nonumber \\
\end{alignat}
Cuts of the type $(1,3,5)$ have the propagator $6$ un-cut. To this propagator we attach the non-trivial
propagator factor as follows:
\begin{alignat}{2}
\begin{picture}(140,60)(0,40)
 \Line(22.5,50)(22.5,70)
 \Line(37.5,50)(37.5,70)
 \Line(70,50)(70,70)
 \Line(110,50)(110,70)
 \Line(30,50)(50,30)
 \Line(50,30)(70,50)
 \Line(110,50)(70,10)
 \Line(70,10)(70,-10)
 \GOval(30,50)(10,12.5)(0){0.5}
 \GCirc(70,50){10}{0.5}
 \GCirc(110,50){10}{0.5}
 \Text(22.5,75)[b]{\small $k_1^+$}
 \Text(37.5,75)[b]{\small $\bar{k}_2^-$}
 \Text(70,75)[b]{\small $k_2^+$}
 \Text(110,75)[b]{\small $k_3^+$}
 \SetColor{Red}
 \SetWidth{2.0}
 \Line(50,30)(70,10)
 \SetWidth{0.5}
 \SetColor{Black}
 \Vertex(50,30){2}
 \Vertex(70,10){2}
\end{picture}
 & \;\; = \;\; \frac{24}{19},
 & 
 \begin{picture}(140,60)(0,40)
 \Line(22.5,50)(22.5,70)
 \Line(37.5,50)(37.5,70)
 \Line(70,50)(70,70)
 \Line(110,50)(110,70)
 \Line(30,50)(50,30)
 \Line(50,30)(70,50)
 \Line(110,50)(70,10)
 \Line(70,10)(70,-10)
 \GOval(30,50)(10,12.5)(0){0.5}
 \GCirc(70,50){10}{0.5}
 \GCirc(110,50){10}{0.5}
 \Text(22.5,75)[b]{\small $k_1^+$}
 \Text(37.5,75)[b]{\small $\bar{k}_2^-$}
 \Text(70,75)[b]{\small $k_2^+$}
 \Text(110,75)[b]{\small $k_3^-$}
 \SetColor{Red}
 \SetWidth{2.0}
 \Line(50,30)(70,10)
 \SetWidth{0.5}
 \SetColor{Black}
 \Vertex(50,30){2}
 \Vertex(70,10){2}
\end{picture}
 & \;\; = \;\; \frac{11}{10}, \nonumber \\
 & & & \nonumber \\
 & & & \nonumber \\
\begin{picture}(140,60)(0,40)
 \Line(22.5,50)(22.5,70)
 \Line(37.5,50)(37.5,70)
 \Line(70,50)(70,70)
 \Line(110,50)(110,70)
 \Line(30,50)(50,30)
 \Line(50,30)(70,50)
 \Line(110,50)(70,10)
 \Line(70,10)(70,-10)
 \GOval(30,50)(10,12.5)(0){0.5}
 \GCirc(70,50){10}{0.5}
 \GCirc(110,50){10}{0.5}
 \Text(22.5,75)[b]{\small $k_1^+$}
 \Text(37.5,75)[b]{\small $\bar{k}_2^+$}
 \Text(70,75)[b]{\small $k_2^-$}
 \Text(110,75)[b]{\small $k_3^+$}
 \SetColor{Red}
 \SetWidth{2.0}
 \Line(50,30)(70,10)
 \SetWidth{0.5}
 \SetColor{Black}
 \Vertex(50,30){2}
 \Vertex(70,10){2}
\end{picture}
 & \;\; = \;\; \frac{370}{361}, 
 & 
 \begin{picture}(140,60)(0,40)
 \Line(22.5,50)(22.5,70)
 \Line(37.5,50)(37.5,70)
 \Line(70,50)(70,70)
 \Line(110,50)(110,70)
 \Line(30,50)(50,30)
 \Line(50,30)(70,50)
 \Line(110,50)(70,10)
 \Line(70,10)(70,-10)
 \GOval(30,50)(10,12.5)(0){0.5}
 \GCirc(70,50){10}{0.5}
 \GCirc(110,50){10}{0.5}
 \Text(22.5,75)[b]{\small $k_1^+$}
 \Text(37.5,75)[b]{\small $\bar{k}_2^+$}
 \Text(70,75)[b]{\small $k_2^-$}
 \Text(110,75)[b]{\small $k_3^-$}
 \SetColor{Red}
 \SetWidth{2.0}
 \Line(50,30)(70,10)
 \SetWidth{0.5}
 \SetColor{Black}
 \Vertex(50,30){2}
 \Vertex(70,10){2}
\end{picture}
 & \;\; = \;\; \frac{24}{19}, \nonumber \\
 & & & \nonumber \\
\end{alignat}
With these vertex and propagator factors we reproduce the combinatorial factor of the underyling chain graph.
}

\section{Diagrams with higher powers of the propagators}
\label{sect:higher_powers}

Let us now discuss loop diagrams with higher powers of the propagators.
These arise from self-energy insertions on internal lines.
An example is shown in fig.~\ref{fig_internal_self_energy}.
These diagrams contribute to the loop amplitude.
These diagrams are characterised by the fact, that at least one propagator occurs to power two or higher.
We may still compute these diagrams within the loop-tree duality method with the help of the general
formula given in eq.~(\ref{general_loop_tree_duality}).
However, this is inconvenient, as this requires the computation of a residue of a function with higher poles.
To see this, let us consider the univariate case.
If $f(z)$ is a function of a complex variable $z$, which has a pole of order $\nu$ at $z_0$, the standard formula for the 
residue at $z_0$ is given
by
\bq
\label{residue_example_one_dim}
 \mathrm{res}\left(f,z_0\right)
 & = &
 \frac{1}{\left(\nu-1\right)!}
 \left.
 \left(\frac{d}{dz}\right)^{\nu-1}
 \left[ \left(z-z_0\right)^\nu f\left(z\right) \right]
 \right|_{z=z_0}.
\eq
We may think of the variable $z$ as being the energy flowing through the raised propagator.
For $\nu > 1$ we have a derivative acting on all $z$-dependent quantities in the diagram.
Although this can be done, it is process-dependent and not very well suited for automation.

\begin{figure}
\begin{center}
\includegraphics[scale=1.0]{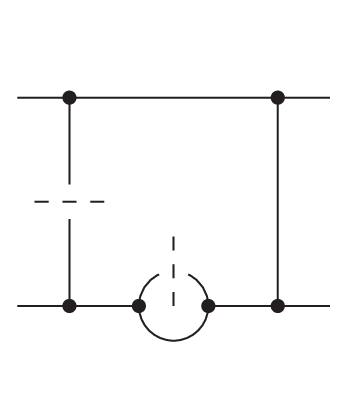}
\hspace*{20mm}
\includegraphics[scale=1.0]{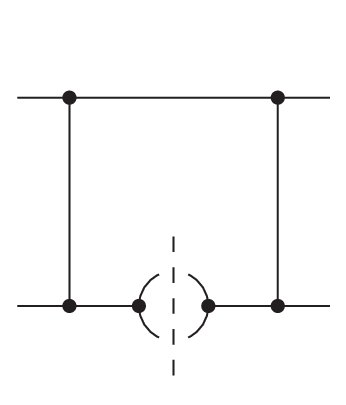}
\hspace*{20mm}
\includegraphics[scale=1.0]{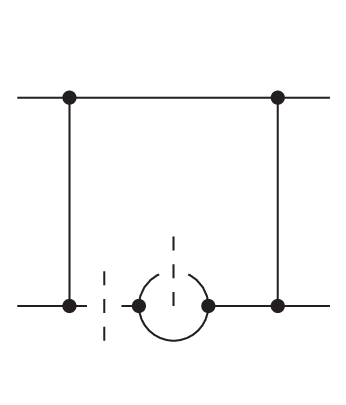}
\end{center}
\caption{
Various cut trees for a two-loop diagram with a self-energy insertion on an internal line.
The corresponding residues for the graphs on the left and in the middle are not problematic and correspond
to residues obtained from single poles.
However, the residue corresponding to the cut shown in the right graph requires the calculation of a residue 
of a function with higher poles.
}
\label{fig_higher_powers}
\end{figure}

In fig.~\ref{fig_higher_powers} we show a selection of $\loopnumber$-fold cuts for the diagram of fig.~\ref{fig_internal_self_energy}.
Please note that not all cuts are problematic.
For example, the cuts shown in the left graph and in the middle graph of fig.~\ref{fig_higher_powers}
are not problematic and correspond to residues obtained from single poles.
However, the residue corresponding to the cut shown in the right graph of fig.~\ref{fig_higher_powers}
requires the calculation of a residue of a function with higher poles.
If we view the right graph of fig.~\ref{fig_higher_powers} as a tree graph, we see that it corresponds
to a diagram with a singular forward limit as discussed 
in eq.~(\ref{condition_forward_higher_power}) in section~\ref{sect:forward_limit}.

In \cite{Baumeister:2019rmh} it was shown that the residues of these cuts cancel with corresponding contributions
from the ultraviolet counterterms for the field renormalisation and the mass renormalisation in the on-shell scheme.
For this reason we included the ultraviolet counterterms from the beginning.
We may therefore simply drop these contributions from the loop amplitude and the counterterms.

Let us discuss the ultraviolet counterterms in more detail.
For all ultraviolet counterterms (field renormalisation, mass renormalisation, coupling renormalisation, etc.)
we use an integral representation.
Let us write 
\bq
 f^{\mathrm{CT}}_{\loopnumber_{\mathrm{CT}},n_{\mathrm{CT}}}\left(\Gamma\right)
\eq
for the integrand of a $\loopnumber_{\mathrm{CT}}$-loop counterterm with $n_{\mathrm{CT}}$ external legs.
A one-loop propagator counter\-term is therefore denoted by $f^{\mathrm{CT}}_{1,2}$,
a one-loop three-valent vertex counterterm by $f^{\mathrm{CT}}_{1,3}$, etc..
{\sw Without loss of generality we may assume that all external momenta of a given counterterm are loop momenta of further loop integrations.
The case, where one or more external momenta of a given counterterm are external momenta of the process under consideration is a simple
specialisation.
}
Let $\alpha \subset \{1,...,\loopnumber\}$ and {\sw $l_{\mathrm{CT}} = | \alpha|$}.
Let further $(q_1,...,q_{n_{\mathrm{CT}}})$ be the set of external momenta for a given counterterm.
With this notation, the integral representation $f^{\mathrm{CT}}_{\loopnumber_{\mathrm{CT}},n_{\mathrm{CT}}}$ has 
the following properties:
\begin{enumerate}
\item the integral 
\bq
\label{integral_repr_CT}
 \int \left( \prod\limits_{j=1}^{\sw l_{\mathrm{CT}}} \frac{d^Dk_{\alpha_j}}{\left(2\pi\right)^D} \right)
 f^{\mathrm{CT}}_{\loopnumber_{\mathrm{CT}},n_{\mathrm{CT}}}
\eq
reproduces the analytic result for the counterterm,
\item the sum of the bare contribution and the counterterm falls off at least 
like ${\mathcal O}(|k_{\alpha_j}|^{-5})$ for $|k_{\alpha_j}|\rightarrow \infty$,
\item the integral representation $f^{\mathrm{CT}}_{\loopnumber_{\mathrm{CT}},n_{\mathrm{CT}}}$ depends only on the spatial components $\vec{q}_1$, ..., $\vec{q}_{n_{\mathrm{CT}}}$,
but not on the energies of $q_1$, ..., $q_{n_{\mathrm{CT}}}$.
\end{enumerate}
We use loop-tree duality also for the integrals involving $f^{\mathrm{CT}}_{\loopnumber_{\mathrm{CT}},n_{\mathrm{CT}}}$.
The last condition ensures that a counterterm integral of the form of eq.~(\ref{integral_repr_CT}) requires exactly 
$\loopnumber_{\mathrm{CT}}$ cuts.
Hence, there is for example no contribution from a propagator counterterm for the cut shown
in the middle diagram of fig.~\ref{fig_higher_powers}.
Of course, there is a contribution from the bare diagram.

In order to avoid higher powers of the propagators from self-energy insertions on internal lines
we choose for the field renormalisation and the mass renormalisation the on-shell scheme.
This allows us to choose integral representations for the propagator counterterms $f^{\mathrm{CT}}_{\loopnumber_{\mathrm{CT}},2}$
(with external momenta $(q,-q)$)
such that
\begin{enumerate}
\setcounter{enumi}{3}
\item 
the sum of all contributing two-point integrands (bare integrands and counterterm integrands) at a given loop order
vanishes quadratically as $q$ goes on-shell.
\end{enumerate}
This condition ensures that there are no contributions from residues related to higher poles.
For example, this condition ensures that the cut shown in the right diagram of fig.~\ref{fig_higher_powers}
gives no contribution, when summed over the bare contribution and the counterterm contribution.

\section{The integrand of the renormalised loop amplitude}
\label{sect:integrand}

In the previous section we introduced
$f^{\mathrm{CT}}_{\loopnumber_{\mathrm{CT}},n_{\mathrm{CT}}}$ as the integrand in $D$-dimensional loop momentum space of a counterterm
of order $\coupling^{2\loopnumber_{\mathrm{CT}}}$ with $n_{\mathrm{CT}}$ external legs.
We may apply loop-tree duality to the integrated counterterm and obtain the integrand in $(D-1)$-dimensional loop momentum
space as the sum over all $\loopnumber_{\mathrm{CT}}$-fold cuts.
Let us denote this expression by
\bq
\label{counterterm_vertex}
 V^{\mathrm{CT}}_{\loopnumber_{\mathrm{CT}},n_{\mathrm{CT}}}.
\eq
We may view {\sw the expression in eq.~(\ref{counterterm_vertex}) } as a new vertex with $n_{\mathrm{CT}}$ external legs and $\loopnumber_{\mathrm{CT}}$ pairs
$(k_{\alpha_j},\bar{k}_{\alpha_j})$, which when sewed together reproduce the integrand $f^{\mathrm{CT}}_{\loopnumber_{\mathrm{CT}},n_{\mathrm{CT}}}$.
\begin{figure}
\begin{center}
\includegraphics[scale=1.0]{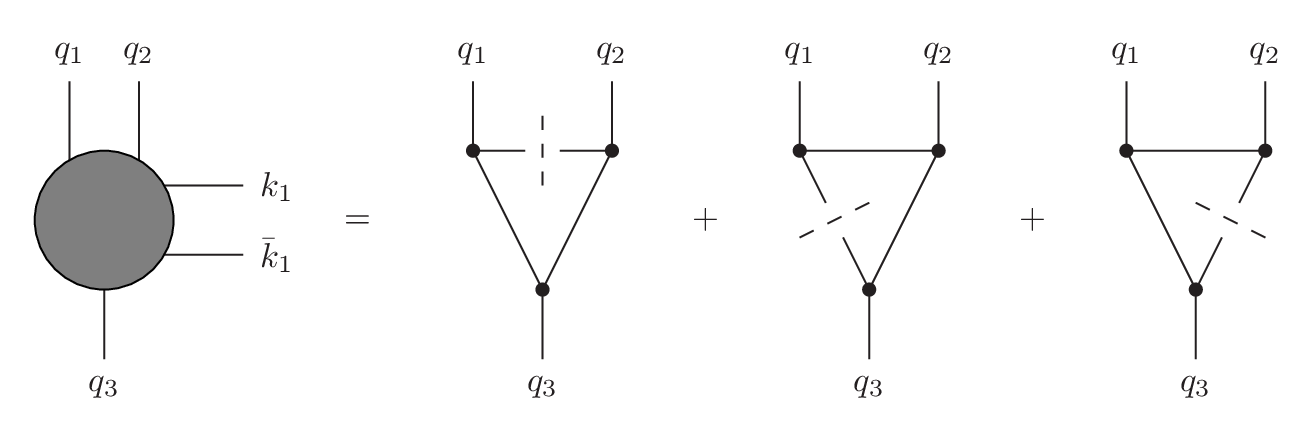}
\end{center}
\caption{
Graphical representation of the integrand of a one-loop counterterm for a three-valent vertex.
Each cut comes with two orientations.
}
\label{fig_counterterms}
\end{figure}
An example is shown in fig.~\ref{fig_counterterms}.
Please note that the vertex $V^{\mathrm{CT}}_{\loopnumber_{\mathrm{CT}},n_{\mathrm{CT}}}$ in fig.~\ref{fig_counterterms}
represents the sum of all cuts shown on the right in fig.~\ref{fig_counterterms}.
{\sw
Note that each cut has its own combinatorial factors $S_{\sigma_{\mathrm{CT}} \alpha_{\mathrm{CT}}}$. The combinatorial factors of a counterterm diagram
need not be identical to the associated bare diagram.
}
Vertices $V^{\mathrm{CT}}_{0,n_{\mathrm{CT}}}$ with $\loopnumber_{\mathrm{CT}}=0$ are the original vertices of the theory.

Having introduced the new vertices $V^{\mathrm{CT}}_{\loopnumber_{\mathrm{CT}},n_{\mathrm{CT}}}$, we may now define 
{\sw a tree-amplitude-like object}
\bq
 {\mathcal A}_{0,n+2\left(\loopnumber-\loopnumber_{\mathrm{CT}}\right),\loopnumber_{\mathrm{CT}}}^{\mathrm{CT}}
\eq
This is {\sw the sum of all tree diagrams} with $(n+2\loopnumber-2\loopnumber_{\mathrm{CT}})$ external legs with vertices consisting of the
original vertices of the theory plus the counterterm vertices $V^{\mathrm{CT}}_{\loopnumber_{\mathrm{CT}},n_{\mathrm{CT}}}$.
{\sw Each diagram is weighted by a combinatorial factor $S_{\sigma \alpha}$}.
The order of all counterterm vertices appearing in ${\mathcal A}_{0,n+2\left(\loopnumber-\loopnumber_{\mathrm{CT}}\right),\loopnumber_{\mathrm{CT}}}^{\mathrm{CT}}$
is $\coupling^{2\loopnumber_{\mathrm{CT}}}$.
{\sw For $\loopnumber_{\mathrm{CT}}=0$ we have
\bq
\label{ampl_no_counterterm}
 {\mathcal A}_{0,n+2\loopnumber,0}^{\mathrm{CT}}
 & = &
 \sum\limits_{\Gamma \in {\mathcal U}_{0,n+2\loopnumber}}
 S_{\sigma \alpha} \;
 f\left(\Gamma\right).
\eq
This differs from the tree-amplitude ${\mathcal A}_{0,n+2\loopnumber,0}$ with momenta $p_1$, ..., $p_n$, $k_1$, ..., $k_l$, $\bar{k}_\loopnumber$, ..., $\bar{k}_\loopnumber$
by the combinatorial factors $S_{\sigma \alpha}$ multiplying each Feynman diagram.
}
The regularised forward limit
\bq
 {\mathcal R}_f {\mathcal A}_{0,n+2\left(\loopnumber-\loopnumber_{\mathrm{CT}}\right),\loopnumber_{\mathrm{CT}}}^{\mathrm{CT}}
\eq
is defined analogously as in section~\ref{sect:forward_limit}.
We call
\bq
{\sw
 {\mathcal B}_{\loopnumber,n}\left(p_1,...,p_n,k_1,...,k_\loopnumber,\bar{k}_1,...,\bar{k}_\loopnumber\right)
}
 & = &
\sum\limits_{\loopnumber_{\mathrm{CT}}=0}^\loopnumber {\mathcal R}_f {\mathcal A}_{0,n+2\left(\loopnumber-\loopnumber_{\mathrm{CT}}\right),\loopnumber_{\mathrm{CT}}}^{\mathrm{CT}}
\eq
the {\sw UV-subtracted regularised forward limit of a weighted sum of tree diagrams, or tree-amplitude-like object for short}.

We now put all the pieces together.
Starting from eq.~(\ref{loop_amplitude_Feynman_diagrams})
\bq
 {\mathcal A}_{\loopnumber,n}\left(p_1,...,p_n\right)
 & = &
 \sum\limits_{\Gamma \in {\mathcal U}_{\loopnumber,n}^{\mathrm{loop}}}
 \frac{\left(-1\right)^{l_{\mathrm{cfl}}(\Gamma)}}{\left|\mathrm{Aut}\left(\Gamma\right)\right|}
 \int \left( \prod\limits_{j=1}^\loopnumber \frac{d^Dk_j}{\left(2\pi\right)^D} \right)
 f\left(\Gamma\right),
\eq
we apply loop-tree duality to all graphs:
\bq
 {\mathcal A}_{\loopnumber,n}\left(p_1,...,p_n\right)
 & = &
{\sw
 \left(-i\right)^\loopnumber
 \sum\limits_{\Gamma \in {\mathcal U}_{\loopnumber,n}^{\mathrm{loop}}}
 \frac{\left(-1\right)^{l_{\mathrm{cfl}}(\Gamma)}}{\left|\mathrm{Aut}\left(\Gamma\right)\right|}
 \sum\limits_{\sigma \in {\mathcal C}_\Gamma}
 \int\limits_{+/-} \left( \prod\limits_{j=1}^\loopnumber \frac{d^{D-1}k_{\sigma_j}}{\left(2\pi\right)^{D-1} 2 \sqrt{\vec{k}_{\sigma_j}^2 + m_{\sigma_j}^2 }} \right)
 S_{\sigma \alpha} \;
 f'\left(\Gamma\right),
 \nonumber
}
\eq
where $f'(\Gamma)$ denotes the integrand without the cut propgators.
We relabel the loop integration momenta $( k_{\sigma_1}, k_{\sigma_2}, ..., k_{\sigma_\loopnumber} )$ to $(k_1,k_2,...,k_\loopnumber)$.
There are $\loopnumber!$ possibilities to do that and we average over all of them.
We exchange summation and integration and obtain
\bq
 {\mathcal A}_{\loopnumber,n}\left(p_1,...,p_n\right)
 & = &
{\sw
 \frac{\left(-i\right)^\loopnumber}{l!}
 \int\limits_{+/-} \left( \prod\limits_{j=1}^\loopnumber \frac{d^{D-1}k_{j}}{\left(2\pi\right)^{D-1} 2 \sqrt{\vec{k}_{j}^2 + m_{j}^2 }} \right)
 \sum\limits_{\Gamma \in {\mathcal U}_{\loopnumber,n}^{\mathrm{loop}}}
 \sum\limits_{\sigma \in {\mathcal C}_\Gamma}
 \sum\limits_{S_l}
 \frac{\left(-1\right)^{l_{\mathrm{cfl}}(\Gamma)}}{\left|\mathrm{Aut}\left(\Gamma\right)\right|}
 S_{\sigma \alpha} \;
 f'\left(\Gamma\right),
 \nonumber 
}
\eq
We recall that we defined in eq.~(\ref{def_forward_backward}) the phase space integration for cut graphs
as an integration over the forward and the backward hyperboloid.
This is equivalent to a sum over both orientations of the momentum flow (or the energy flow).
Making this sum explicit we recognise that the four sums make up the set
${\mathcal U}_{\loopnumber,n}^{\loopnumber-\mathrm{marked}, \mathrm{\; non-singular}}$.
We therefore have
\bq
\lefteqn{
 {\mathcal A}_{\loopnumber,n}\left(p_1,...,p_n\right)
 = } & &
 \nonumber \\
 & &
{\sw
 \frac{\left(-i\right)^\loopnumber}{l!}
 \int \left( \prod\limits_{j=1}^\loopnumber \frac{d^{D-1}k_{j}}{\left(2\pi\right)^{D-1} 2 \sqrt{\vec{k}_{j}^2 + m_{j}^2 }} \right)
 \sum\limits_{\Gamma \in {\mathcal U}_{\loopnumber,n}^{\loopnumber-\mathrm{marked}, \mathrm{\; non-singular}}}
 \frac{\left(-1\right)^{l_{\mathrm{cfl}}(\Gamma)}}{\left|\mathrm{Aut}\left(\Gamma\right)\right|}
 \; S_{\sigma \alpha} \;
 f'\left(\Gamma\right).
 \nonumber
}
\eq
Let us emphasise that for marked graphs the phase space integration does not include a sum over the forward and the 
backward hyperboloids. This sum corresponds to the two possible orientations of each marked line 
and is included in the set ${\mathcal U}_{\loopnumber,n}^{\loopnumber-\mathrm{marked}}$.

We then use eq.~(\ref{graph_summation}) to exchange the summation over marked graphs with the summation over sewed graphs.
There are no singular propagators in the summation: Singular propagators due to higher powers of some propagator 
(i.e. due to self-energy corrections on internal lines) 
cancel in the combination of bare and counterterm contributions in the on-shell scheme.
Singular propagators from self-energy corrections on external lines are absent from the beginning due to the LSZ reduction
formula.
Finally, in theories where all field have a vanishing vacuum expectation values we may neglect contributions from tadpoles,
hence there are no singular propagators due to tadpoles.
We therefore recognise the sum over sewed graphs as the regularised forward limit and obtain 
\bq
\label{main_result}
 {\mathcal A}_{\loopnumber,n}\left(p_1,...,p_n\right)
 & = &
{\sw
 \frac{\left(-i\right)^\loopnumber}{\loopnumber!}
 \int \left( \prod\limits_{j=1}^\loopnumber \frac{d^{D-1}k_{j}}{\left(2\pi\right)^{D-1} 2 \sqrt{\vec{k}_{j}^2 + m_{j}^2 }} \right)
  \sum\limits_{\loopnumber_{\mathrm{CT}}=0}^\loopnumber {\mathcal R}_f {\mathcal A}_{0,n+2\left(\loopnumber-\loopnumber_{\mathrm{CT}}\right),\loopnumber_{\mathrm{CT}}}^{\mathrm{CT}}
}
 \nonumber \\
 & = &
{\sw 
 \frac{\left(-i\right)^\loopnumber}{\loopnumber!}
 \int \left( \prod\limits_{j=1}^\loopnumber \frac{d^{D-1}k_{j}}{\left(2\pi\right)^{D-1} 2 \sqrt{\vec{k}_{j}^2 + m_{j}^2 }} \right)
  {\mathcal B}_{\loopnumber,n}.
}
\eq
{\sw Eq.~(\ref{main_result}) is the main results of this paper.
This equation expresses} the renormalised loop amplitude as a phase space integral of the regularised forward limit
of a {\sw tree-amplitude-like object ${\mathcal B}_{l,n}$}.
The virtue of {\sw this formula lies in the fact, that it does not} refer to Feynman diagrams.

The propagators in the integrand of {\sw eq.~(\ref{main_result})} 
have a modified (``dual'') $i\delta$-prescrip\-tion, as discussed in section~\ref{sect:loop_amplitudes}.
The dual $i\delta$-prescription is relevant for non-pinch singularities in the phase space integration
in {\sw eq.~(\ref{main_result}).}
The dual $i\delta$-prescription dictates into which direction the contour should be deformed to avoid
non-pinch singularities.

There are a few straightforward generalisations of {\sw eq.~(\ref{main_result}):}
In theories with several particle species we include a sum over all flavours for the $(2\loopnumber)$ additional external particles.
In theories with spin we include a sum over (physical and unphysical) polarisations according to
eq.~(\ref{pol_sum_fermions}) and eq.~(\ref{pol_sum_boson}) for the $(2\loopnumber)$ additional external particles.
In gauge theories we include ghosts and anti-ghosts among the $(2\loopnumber)$ additional external particles.

\section{Recurrence relations}
\label{sect:recurrence_relations}

Tree amplitudes are efficiently computed using recurrence relations \cite{Berends:1987me,Kosower:1989xy,Draggiotis:2002hm,Weinzierl:2005dd,Dinsdale:2006sq,Duhr:2006iq}.
{\sw In this section we discuss the required modifications for the tree-amplitude-like objects ${\mathcal B}_{l,n}$
introduced in the previous section.
}
Let us first review the algorithm for a tree amplitude ${\mathcal A}_{0,n}$.
For simplicity, we focus on $\phi^3$-theory.
The recursive algorithm is based on off-shell currents ${\mathcal J}_{0,j}(q_1,...,q_j)$.
An off-shell current is an object with $j$ external on-shell legs with
momenta $q_1$, ..., $q_j$ and one additional off-shell leg $q_{j+1}$, satisfying momentum conservation
\bq
 q_1 + ... + q_j + q_{j+1} & = & 0.
\eq
The recursive algorithm proceeds as follows:
\begin{enumerate}
\item Initialisation: Set 
\bq
 {\mathcal J}_{0,1}\left(q_i\right) & = & 1,
 \;\;\;\;\;\; i \in \{1,...,j-1\}.
\eq
In theories with spin the one-currents ${\mathcal J}_{0,1}$ are given by the external polarisations, for example
by polarisation vectors for gauge bosons and by spinors for spin $1/2$ fermions.
\item Recursion: Let $\gamma$ be a subset of $\{q_1,...,q_{j-1}\}$ and $\alpha$ and $\beta$ subsets of $\gamma$
with
\bq
\label{condition_subsets}
 \alpha \cup \beta \; = \; \gamma,
 & &
 \alpha \cap \beta \; = \; \emptyset.
\eq
Let $i=|\gamma|$, $a=|\alpha|$ and
\bq
 Q_1 \; = \; \sum\limits_{q \in \alpha} q,
 \;\;\;
 Q_2 \; = \; \sum\limits_{q \in \beta} q,
 \;\;\;
 Q_3 \; = \; -Q_1-Q_2.
\eq
Set
\bq
\label{off_shell_recursion}
 {\mathcal J}_{0,i}^{\mathrm{amputated}}\left(\gamma\right) 
 & = & 
 \sum\limits_{\alpha,\beta} 
 V\left(Q_1,Q_2,Q_3\right)
 {\mathcal J}_{0,a}\left(\alpha\right)
 {\mathcal J}_{0,i-a}\left(\beta\right),
 \nonumber \\
 {\mathcal J}_{0,i}\left(\gamma\right) 
 & = & 
 P\left(Q_3,-Q_3\right)
 {\mathcal J}_{0,i}^{\mathrm{amputated}}\left(\gamma\right),
\eq
where $V(q_1,q_2,q_3)$ denotes the vertex and $P(q_3,-q_3)$ the propagator.
The sum is over all subsets $\alpha$, $\beta$ of $\gamma$ satisfying eq.~(\ref{condition_subsets}).
\begin{figure}
\begin{center}
\includegraphics[scale=1.0]{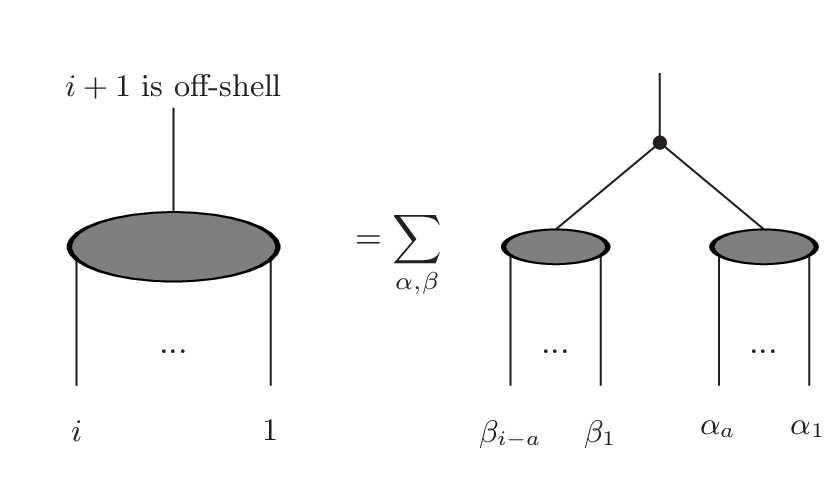}
\end{center}
\caption{
The recurrence relation for the off-shell current in $\phi^3$-theory: The current ${\mathcal J}_{0,i}$
is given as a sum over all sub-currents ${\mathcal J}_{0,|\alpha|}$ and ${\mathcal J}_{0,|\beta|}$
contracted into the three-valent vertex.
The sets $\alpha$ and $\beta$ satisfy $\alpha \cup \beta = \{1,...,i\}$ and $\alpha \cap \beta = \emptyset$.
}
\label{fig_recurrence_relation}
\end{figure}
This is shown schematically in fig.~\ref{fig_recurrence_relation}.

In theories with several vertices, possibly with higher valency, eq.~(\ref{off_shell_recursion})
includes a sum over all allowed vertices. For vertices with higher valency, the subset $\gamma$ is
partitioned into more than two subsets.
\item Amplitude:
The amplitude is given by
\bq
 {\mathcal A}_{0,n}\left(p_1,...,p_n\right)
 & = &
 {\mathcal J}_{0,n-1}^{\mathrm{amputated}}\left(p_1,...,p_{n-1}\right). 
\eq
In theories with spin the amplitude ${\mathcal A}_{0,n}$ 
is given by the contraction of ${\mathcal J}_{0,n-1}^{\mathrm{amputated}}$ with ${\mathcal J}_{0,1}(p_n)$.
\end{enumerate}
In order to compute with this algorithm the UV-subtracted regularised forward limit of 
\linebreak ${\mathcal A}_{0,n+2\left(\loopnumber-\loopnumber_{\mathrm{CT}}\right),\loopnumber_{\mathrm{CT}}}^{\mathrm{CT}}$  we have to make the following {\sw three} modifications:
First, we enlarge the set of vertices and include the
counterterm vertices defined in eq.~(\ref{counterterm_vertex}).
Secondly, we exclude terms which have singular propagators.
{\sw Thirdly, we include combinatorial weight factors for vertices and propagators as discussed in section~\ref{sect:symmetry_factors}.}
We denote by ${\mathcal J}_{\loopnumber_{\mathrm{CT}},j}(q_1,...,q_j)$ the off-shell current with $j$ on-shell legs with
momenta $q_1$, ..., $q_j$ and containing counterterm vertices of order $\coupling^{2\loopnumber_{\mathrm{CT}}}$.
The momenta $q_1$, ..., $q_j$ are a subset of
\bq
 \left\{ p_1, ..., p_n, k_1, ..., k_\loopnumber, \bar{k}_1, ..., \bar{k}_\loopnumber \right\}.
\eq
The off-shell current ${\mathcal J}_{\loopnumber_{\mathrm{CT}},j}(q_1,...,q_j)$ depends in addition on $\loopnumber_{\mathrm{CT}}$ pairs $(k_a,\bar{k}_a)$ with
$k_a, \bar{k}_a \notin \{q_1,...,q_j\}$ through the counterterm vertices.
{\sw The inclusion of the combinatorial factors for the vertices is unproblematic.
The inclusion of the combinatorial factors for the propagators is more tricky, as the combinatorial factors
for the propagators are only known after the off-shell current is contracted into another vertex.
For this reason we treat off-shell currents, which potentially may lead to non-trivial combinatorial factors
as currents of different flavour and combine those only (with the correct combinatorial factors) when they are contracted into the next vertex.
}
Our algorithm for $\phi^3$-theory is given by
\begin{enumerate}
\item Initialisation: Set 
\bq
 {\mathcal J}_{0,1}\left(q_i\right) & = & 1,
 \;\;\;\;\;\; i \in \{1,...,j-1\},
 \nonumber \\
 {\mathcal J}_{k,1}\left(q_i\right) & = & 0,
 \;\;\;\;\;\; k \ge 1,
\eq
\item Recursion: Let $\gamma$ be a subset of $\{q_1,...,q_{j-1}\}$ and $\alpha$ and $\beta$ subsets of $\gamma$
with
\bq
\label{condition_subsets_general}
 \alpha \cup \beta \; = \; \gamma,
 & &
 \alpha \cap \beta \; = \; \emptyset.
\eq
Let $i=|\gamma|$, $a=|\alpha|$ and
\bq
 Q_1 \; = \; \sum\limits_{q \in \alpha} q,
 \;\;\;
 Q_2 \; = \; \sum\limits_{q \in \beta} q,
 \;\;\;
 Q_3 \; = \; -Q_1-Q_2.
\eq
Set
\bq
\label{off_shell_recursion_general}
{\sw
 {\mathcal J}_{\loopnumber_{\mathrm{CT}},i,f}^{\mathrm{amputated}}\left(\gamma\right) 
}
 & = & 
{\sw
 \sum\limits_{\loopnumber_1+\loopnumber_2+\loopnumber_3=\loopnumber_{\mathrm{CT}}}
 \;\;
 \sum\limits_{\alpha,\beta}{}'
 \;
 \sum\limits_{f_1,f_2}
 \; 
 C_V
 V^{\mathrm{CT}}_{\loopnumber_3,3}\left(Q_1,Q_2,Q_3\right)
 C_\alpha
 {\mathcal J}_{\loopnumber_1,a,f_1}\left(\alpha\right)
 C_\beta
 {\mathcal J}_{\loopnumber_2,i-a,f_2}\left(\beta\right),
}
 \nonumber \\
\eq
where {\sw the primed sum} is over all subsets $\alpha$, $\beta$ of $\gamma$ satisfying eq.~(\ref{condition_subsets_general})
{\sw and selecting only those contributions, which will give the same combinatorial factor for the propagator.
The different possibilities for the combinatorial factor for the propagator are indexed by $f$.
$C_V$ denotes the combinatorial factor for the vertex, $C_\alpha$ and $C_\beta$ the combinatorial factors for the
propagators for the sub-currents ${\mathcal J}_{\loopnumber_1,a,f_1}\left(\alpha\right)$ and
${\mathcal J}_{\loopnumber_2,i-a,f_2}\left(\beta\right)$, which can be determined at this stage.
}

If the momentum $(-Q_3)$ is of the form as in eq.~(\ref{condition_forward_on_shell}), (\ref{condition_forward_higher_power}) or (\ref{condition_forward_zero_momentum}) set
\bq
{\sw
 {\mathcal J}_{\loopnumber_{\mathrm{CT}},i,f}\left(\gamma\right) 
}
 & = & 
 0,
\eq
otherwise set
\bq
{\sw
 {\mathcal J}_{\loopnumber_{\mathrm{CT}},i,f}\left(\gamma\right) 
}
 & = & 
 P\left(Q_3,-Q_3\right)
{\sw
 {\mathcal J}_{\loopnumber_{\mathrm{CT}},i,f}^{\mathrm{amputated}}\left(\gamma\right)
}
 \nonumber \\
 & &
 +
 \sum\limits_{\loopnumber_1+\loopnumber_2=\loopnumber_{\mathrm{CT}}}
 P\left(Q_3,-Q_3\right)
 V^{\mathrm{CT}}_{\loopnumber_2,2}\left(Q_3,-Q_3\right)
 P\left(Q_3,-Q_3\right)
{\sw
 {\mathcal J}_{\loopnumber_1,i,f}^{\mathrm{amputated}}\left(\gamma\right).
}
\;\;
\eq
\item The regularised forward limit is given by
\bq
\lefteqn{
 {\mathcal R}_f {\mathcal A}_{0,n+2\left(\loopnumber-\loopnumber_{\mathrm{CT}}\right),\loopnumber_{\mathrm{CT}}}^{\mathrm{CT}}\left(p_1,...,p_n,k_{\alpha_1},...,k_{\alpha_{\loopnumber-\loopnumber_{\mathrm{CT}}}},\bar{k}_{\alpha_1},...,\bar{k}_{\alpha_{\loopnumber-\loopnumber_{\mathrm{CT}}}}\right)
 = } & & \nonumber \\
 & &
 {\mathcal J}_{\loopnumber_{\mathrm{CT}},n-1+2\left(\loopnumber-\loopnumber_{\mathrm{CT}}\right)}^{\mathrm{amputated}}\left(p_1,...,p_{n-1},k_{\alpha_1},...,k_{\alpha_{\loopnumber-\loopnumber_{\mathrm{CT}}}},\bar{k}_{\alpha_1},...,\bar{k}_{\alpha_{\loopnumber-\loopnumber_{\mathrm{CT}}}}\right). 
\eq
\end{enumerate}
This algorithms allows the computation of the UV-subtracted regularised forward limit within $\phi^3$-theory.
In this theory, the only counterterms are two-point counterterms or three-point counterterms.
The algorithm is straightforwardly extended to more general quantum field theories with additional vertices (possibly with higher
valency) and with spins, following the remarks given in the algorithm for the computation of 
${\mathcal A}_{0,n}$.

{\sw
Let us add a technical comment relevant to three-loops and beyond: The combinatorial factors given in 
eq.~(\ref{comb_factor_second_prop}) should only be applied to the second chain with a given loop momentum.
This can be realised by an array of flags, one for each chain. These flags are set to a specific value
after the corresponding current has been calculated for the first time.
}

\section{Fields with non-vanishing vacuum expectation values}
\label{sect:vev}

In the main part of this paper we focused on theories where all fields have a vanishing vacuum expectation value.
This excludes the Higgs sector of the Standard Model, where the Higgs field has a non-vanishing vacuum expectation value.
In this short paragraph we comment on the extension of our results towards the Standard Model.
We have to discuss the treatment of tadpoles, where the tadpole is connected to the rest of the diagram through a 
Higgs propagator. The momentum flow through this propagator is zero, but the Higgs particle has a non-zero mass,
therefore the propagator is non-singular.
We may therefore keep these contributions (the tadpole and the associated UV counterterm as shown in fig.~\ref{fig_tadpole}).
Thus we would define the regularised forward limit in such a way that it includes these contributions.
We modify eq.~(\ref{condition_forward_zero_momentum}) and allow diagrams  
with an internal edge
\bq
 \sum\limits_{a \in \alpha} \left( k_a + \bar{k}_a \right),
\eq
if this edge corresponds to a propagating Higgs particle.

Up to now we did not make any reference to any particular renormalisation scheme in the Higgs sector.
In refs.~\cite{Ross:1973fp,Bohm:1986rj} the renormalisation of the Standard Model is discussed.
In particular, it is convenient to renormalise the vacuum expectation value of the Higgs field such that
it corresponds to the physical vacuum expectation value of the interacting theory.
This renormalisation condition translates to the condition that the tadpole contributions vanish and we may simply omit them.
If this renormalisation condition is imposed, there are no modifications of our result: As in the unbroken case, we do not
include tadpoles in the regularised forward limit.

\section{Checks}
\label{sect:checks}

{\sw
We have checked the basic formula for loop-tree duality eq.~(\ref{loop_tree_duality_single_poles})
together with the explicit values of the combinatorial factors $S_{\sigma \alpha}$ given in
eq.~(\ref{combinatorial_factors_I}) and eq.~(\ref{combinatorial_factors_II}) for a variety of graphs up to three loops.
This can be done in $D=1$ space-time dimensions.
In this case, no integration is left on the right-hand side of eq.~(\ref{loop_tree_duality_single_poles}).
On the other hand, we may easily evaluate the loop integral on the left-hand side for imaginary masses
by Monte Carlo integration. We found for all graphs complete agreement.
Let us give some examples:
Our starting point in $D=1$ space-time dimensions is
\bq
 I & = &
 \int \left( \prod\limits_{j=1}^\loopnumber \frac{dk_j}{2\pi} \right)
 \frac{1}{\prod\limits_{e_j \in E_\Gamma} D_j },
\eq
where $D_j=k_j^2-m_j^2$.
We use the numerical values
\begin{align}
 m_1 & = - 11i,
 &
 m_2 & = - 13i,
 &
 m_3 & = - 17i,
 &
 m_4 & = - 23i,
 \\
 m_5 & = - 31i,
 &
 m_6 & = - 43i,
 &
 m_7 & = - 47i.
\end{align}
For the chain graphs of fig.~(\ref{fig_21}) we find
\bq
\mbox{One-loop:}
 & &
 I^{\mathrm{LT}} \; = \; 
 \frac{1}{22} 
 \; \approx \;
 4.54545 \cdot 10^{-2},
\nonumber \\
 & &
 I^{\mathrm{MC}} 
 \; = \; 
 \left( 4.54545 \pm 0.00007 \right) \cdot 10^{-2},
\nonumber \\
\mbox{Two-loop:}
 & &
 I^{\mathrm{LT}} \; = \;
 \frac{1}{398684}
 \; \approx \;
 2.50825 \cdot 10^{-6},
\nonumber \\
 & &
 I^{\mathrm{MC}} 
 \; = \; 
 \left( 2.5083 \pm 0.0001 \right) \cdot 10^{-6},
\nonumber \\
\mbox{Three-loop:}
 & &
 I^{\mathrm{LT}} \; = \; 
 \frac{3264791}{253676278437997615200}
 \; \approx \;
 1.28699 \cdot 10^{-14},
\nonumber \\
 & &
 I^{\mathrm{MC}} 
 \; = \;
 \left( 1.28699 \pm 0.00008 \right) \cdot 10^{-14}.
\eq
For the two chain graphs of fig.~(\ref{fig_24}) we find
\bq
\mbox{Five propagator graph:}
 & &
 I^{\mathrm{LT}} \; = \; 
 \frac{19}{653441364576}
 \; \approx \;
 2.90768 \cdot 10^{-11},
\nonumber \\
 & &
 I^{\mathrm{MC}} 
 \; = \; 
 \left( 2.9077 \pm 0.0002 \right) \cdot 10^{-11},
 \nonumber \\
\mbox{Four propagator graph:}
 & &
 I^{\mathrm{LT}} \; = \; 
 \frac{1}{28627456}
 \; \approx \;
 3.49315 \cdot 10^{-8},
\nonumber \\
 & &
 I^{\mathrm{MC}} \; = \; 
 \left( 3.4932 \pm 0.0003 \right) \cdot 10^{-8}.
\eq
External momenta do not change the combinatorial factors.
\begin{figure}
\begin{center}
\includegraphics[scale=1.0]{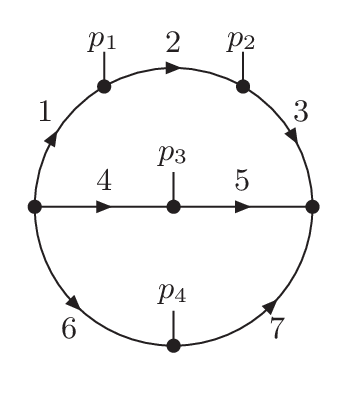}
\hspace*{20mm}
\includegraphics[scale=1.0]{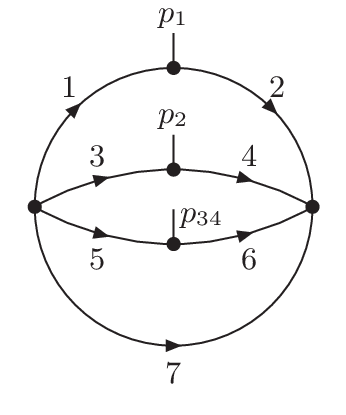}
\end{center}
\caption{
The two-loop non-planar double box (left) and a three-loop vertex graph (right).
All external momenta are outgoing. The notation $p_{34}=p_3+p_4$ is used.
}
\label{fig_28}
\end{figure}
Two non-trivial examples of graphs with external momenta are shown in fig.~(\ref{fig_28}).
With
\begin{align}
 p_1 & = 1,
 & 
 p_2 & = 3,
 & 
 p_3 & = 5,
\end{align}
and $p_4=-p_1-p_2-p_3$
we obtain for these graphs
\bq
\mbox{Non-planar double box:}
 & &
 I^{\mathrm{LT}} 
 \; \approx \;
 9.50190 \cdot 10^{-19},
\nonumber \\
 & &
 I^{\mathrm{MC}} \; = \;
 \left( 9.504 \pm 0.005 \right) \cdot 10^{-19},
 \nonumber \\
\mbox{Three-loop vertex:}
 & &
 I^{\mathrm{LT}}  
 \; \approx \;
 7.92589 \cdot 10^{-18}, 
\nonumber \\
 & &
 I^{\mathrm{MC}} \; = \; 
 \left( 7.928 \pm 0.007 \right) \cdot 10^{-18}.
\eq
Please note that the verification of loop-tree duality in $D=1$ space-time dimensions 
implies a verification in arbitrary space-time dimensions.
We may always substitute the squared internal masses by
\bq
 m_j^2 & \rightarrow & m_j^2 + \vec{q}_j^2,
\eq
where $\vec{q}_j$ denotes a $(D-1)$-dimensional spatial momentum, depending on the 
$(D-1)$-dimen\-sional spatial loop momenta $\vec{k}_1$, $\dots$, $\vec{k}_l$ 
and possibly on the external spatial momenta $p_1$, $\dots$, $p_n$.
The loop integral in $D$ space-time dimensions is given by the integral over all spatial loop momenta.
The verification in one space-time dimension implies the equality of integrands of the spatial integrations.
The equality of the integrands implies of course the equality of the integrals.

We may check this for integrals, which neither require subtraction terms nor contour deformation.
Our main interest is $D=4$.
Examples are provided by Feynman integrals, which are ultraviolet- and infrared-finite 
and which are evaluated in the Euclidean region (i.e. all Lorentz invariants non-positive).
The former condition (ultraviolet- and infrared-finiteness) is a necessary condition that no subtraction terms are required,
the latter condition (Euclidean region) ensures that no contour deformation is required.
\begin{figure}
\begin{center}
\includegraphics[scale=1.0]{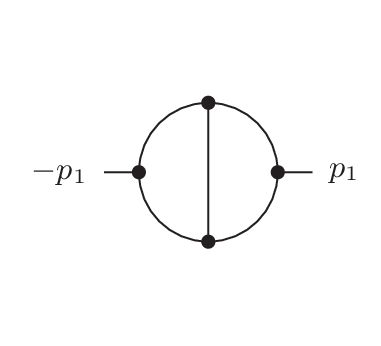}
\hspace*{5mm}
\includegraphics[scale=1.0]{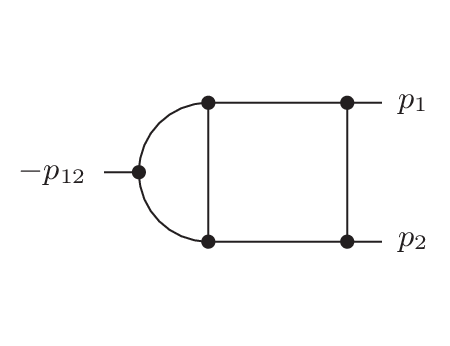}
\hspace*{8mm}
\includegraphics[scale=1.0]{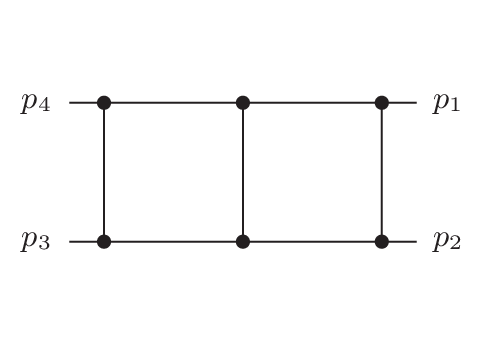}
\end{center}
\caption{
Finite integrals in four space-time dimensions. All external momenta are massive, all internal lines are massless.
}
\label{fig_30}
\end{figure}
Fig.~\ref{fig_30} show a few two-loop examples, where analytic results are available \cite{Usyukina:1993ch}.
All external momenta are massive, all internal lines are massless.
The normalisation of the integrals is
\bq
I  & = &
 \int \prod\limits_{r=1}^{l} \frac{d^4k_r}{(2\pi)^4}
 \;
 \prod\limits_{j=1}^{n} D_j,
\eq
where $D_j=k_j^2$.
As external momenta we use
\begin{align}
 p_1 & = \left(1, 7, 11, 13 \right),
 & 
 p_2 & = \left( 3, 17, 19, 23 \right),
 & 
 p_3 & = \left( 5, 29, 31, 37 \right).
\end{align}
The Lorentz invariants are then
\bq
 & &
 p_1^2 = -338,
 \;\;\;\;\;\;
 p_2^2 = -1170,
 \;\;\;\;\;\;
 p_3^2 = -3146,
 \;\;\;\;\;\;
 p_4^2 = -11778,
 \\
 & &
 s = \left(p_1+p_2\right)^2 = -2756,
 \;\;\;\;\;\;
 t = \left(p_2+p_3\right)^2 = -8152,
 \;\;\;\;\;\;
 u = \left(p_1+p_3\right)^2 = -5524.
 \nonumber 
\eq
We find good agreement between the loop-tree duality representation (which we evaluate with Monte Carlo techniques) and the analytic result:
\bq
\mbox{Two-point function:}
 & &
 I^{\mathrm{LT}} 
 \; \approx \;
 \left( 8.549 \pm 0.007 \right) \cdot 10^{-7},
\nonumber \\
 & &
 I^{\mathrm{analytic}} \; = \;
 8.557 \cdot 10^{-7},
 \nonumber \\
\mbox{Vertex:}
 & &
 I^{\mathrm{LT}} 
 \; \approx \;
 \left( -1.192 \pm 0.001 \right) \cdot 10^{-10},
\nonumber \\
 & &
 I^{\mathrm{analytic}} \; = \;
 -1.193 \cdot 10^{-10},
 \nonumber \\
\mbox{Double box:}
 & &
 I^{\mathrm{LT}} 
 \; \approx \;
 \left( 1.680 \pm 0.001 \right) \cdot 10^{-14},
\nonumber \\
 & &
 I^{\mathrm{analytic}} \; = \;
 1.680 \cdot 10^{-14}.
\eq
In order to check that the sum of graphs is correct, we proceeded as follows:
We consider in $\phi^3$-theory scattering processes up to three loops with $n + 2 l < 10$.

On the one hand, we use \verb|QGRAF| \cite{Nogueira:1993ex} to generate all (bare) loop graphs.
For each loop graph, we obtain the set of cut graphs.
For each cut of a single propagator we sum over the two possibilities to label the half-edges by $(k,\bar{k})$ 
or $(\bar{k},k)$ (corresponding to the two possible orientations).
For multi-loop graphs ($\loopnumber \ge 2$) we average over the $\loopnumber !$ possibilities of assigning
$k_1,...,k_\loopnumber$  to the half-edges.
We remove graphs which correspond to residues of functions with higher poles.
This gives a list of cut graph, together with a numerical factor, given by the product of the usual
symmetry factor and $1/\loopnumber!$.
The combinatorial factor is taken care by the vertex and propagator factors. 
Identical graphs are combined by adding their numerical factors.
We have verified that this cancels all symmetry factors, resulting in a numerical factor $1/\loopnumber!$ for all graphs.

On the other hand, we generate a list of cut graphs from the off-shell recurrence relations, including
the factor $1/l!$ in eq.~(\ref{main_result}).
We neglect UV-counterterms.
For $n+2l=9$ the lists consist of ${\mathcal O}(10^5)$ graphs.
Comparing the two lists of graphs (with the help of a computer program), we find agreement including all numerical factors.

Dressing up each graph with a graph-dependent numerator will not change the combinatorics.
Graphs with vertices of valency four or higher can always be written as graphs with three-valent vertices
only and numerators, which cancel the extra propagators. 
The basic formula of loop-tree duality is not affected by numerators.
In fact, we discussed in section~\ref{sect:loop_amplitudes} the general case with numerators.
This implies that our check of $\phi^3$-theory tests also the essential combinatorial parts
of any other theory, including Yang-Mills theory.

Let us add one technical remark: In gauge theories the use of Feynman gauge is the most natural
choice for our purposes, as it avoids the introduction of higher or spurious poles in the propagator.
When cutting a gauge boson line, we have to replace the numerator $(-g_{\mu\nu})$ by a polarisation sum.
This cannot be done with just physical polarisation, but we need to introduce unphysical polarisations
as discussed in eq.~(\ref{pol_sum_boson}).
If we restrict our attention just to the (bare) loop amplitude, the contribution from the unphysical 
polarisations does not drop out: This can be seen already at one-loop by looking at the regions giving
rise to collinear singularities. This requires two adjacent loop propagators to go on-shell, where one of the
two on-shell propagators carries an unphysical polarisation \cite{Nagy:2003qn}.
In contrast, the collinear singularity in the real emission contribution has two collinear particles,
where both particles have physical polarisations.
From the mismatch of the polarisations it is clear that these two contributions can never cancel locally.
The solution of this paradox is as follows: A local cancellation of collinear singularities is achieved,
if an integral representation of the field renormalisation constants is included.
Then the longitudinal part of the collinear singularity from the loop amplitude cancels locally
with the longitudinal part of the collinear singularity of the integral representation 
of the field renormalisation constant, while the
the transverse part of the collinear singularity from the real emission cancels locally
with the transverse part of the collinear singularity of the integral representation 
of the field renormalisation constant \cite{Seth:2016hmv}.
This demonstrates that unphysical polarisations and ghosts are required within the loop-tree duality approach
in Feynman gauge.
}

\section{Conclusions and outlook}
\label{sect:conclusions}

In this paper we showed that the integrand of a renormalised loop amplitude can be related within loop-tree duality 
to the regularised forward limit
of a UV-subtracted {\sw tree-amplitude-like object}.
This nice form is achieved if field renormalisation and mass renormalisation are performed in the on-shell scheme.
The use of the on-shell renormalisation scheme for these two quantities eliminates contributions from residues underlying higher poles.
Our final result gives the integrand in terms of {\sw tree-amplitude-like objects}, not individual Feynman diagrams.
This has several advantages:
First of all, it allows for an efficient computation: As ordinary tree amplitudes, the UV-subtracted regularised forward limit
can be computed efficiently with recurrence relations.
Secondly, our definition of the loop integrand naturally includes a global definition of the loop momenta.
We expect this representation to have particular nice properties with respect to a local cancellation of infrared singularities
with the corresponding real emission parts.

Let us outline how the result of this paper fits into the bigger goal of numerical higher-order computations:
Within loop-tree duality all contributions at a given order in perturbation theory
(from the purely virtual contribution to the purely real emission contribution) 
live on spaces of the same dimension.
It is common practice to use for the Monte Carlo integration of the real emission part a multi-channel approach, corresponding
to the individual infrared limits.
In each channel we may then in the next step set-up mappings between the real emission part and the parts involving loops.
At NLO, these mappings can be found in \cite{Sborlini:2016gbr,Seth:2016hmv}.
With the mappings at hand, one then proves the local cancellation of infrared singularities.
This treats infrared singularities, where a cancellation occurs between real and virtual parts.
In a final step, and before embarking on a numerical Monte Carlo integration we must also treat singularities
in the real subspace of the virtual part. These singularities are handled with contour deformation.
Algorithms to construct a suitable contour are already available, even at higher loops \cite{Gong:2008ww,Becker:2012nk,Becker:2012bi,Buchta:2015wna,Capatti:2019edf}.

\subsection*{Acknowledgements}

We would like to thank H. Spiesberger for useful discussions.

This work has been supported by the 
Cluster of Excellence ``Precision Physics, Fundamental Interactions, and Structure of Matter'' 
(PRISMA+ EXC 2118/1) funded by the German Research Foundation (DFG) 
within the German Excellence Strategy (Project ID 39083149).

{\footnotesize
\bibliography{/home/stefanw/notes/biblio}
\bibliographystyle{/home/stefanw/latex-style/h-physrev5}
}

\end{document}